\documentclass[twocolumn,floatfix,longbibliography]{revtex4-2}  
\usepackage{graphics,epsfig,amsfonts,amssymb,amsmath,xcolor}  
\usepackage[normalem]{ulem}
\begin{document}
\title{
 Mott correlations in ABC graphene trilayer aligned with hBN}

\author{
M.J. Calder\'on$^1$, 
A. Camjayi$^2$, 
E. Bascones$^1$}
\affiliation{$^1$ Instituto de Ciencia de Materiales de Madrid (ICMM). Consejo Superior de Investigaciones Cient\'ificas (CSIC), Sor Juana In\'es de la Cruz 3, 28049 Madrid (Spain).\\
$^2$ Ciclo B\'asico Com\'un, Universidad de Buenos Aires and IFIBA, Conicet, Pabell\'on 1, Ciudad Universitaria, Buenos Aires 1428 CABA, Argentina}
\date{\today}

\begin{abstract}
The nature of the correlated phases found in some graphene heterostructures is under debate. We use dynamical mean-field theory (DMFT) to analyze the effect of local correlations close to half-filling on one of such systems, the ABC trilayer graphene aligned with hexagonal boron nitride (ABC/hBN), which presents a moir\'e superlattice. This system has shown insulating phases at integer fillings of the moir\'e lattice, precisely the fillings at which a sufficiently strong Coulomb interaction (U$_{\rm Mott}$) may produce a metal-insulator Mott transition. Our calculations show that the electronic states are strongly affected by a significant spectral weight transfer at interactions  with magnitudes expected to be relevant in experiments.  This effect, which emerges at interactions considerably smaller than U$_{\rm Mott}$ and does not require symmetry breaking, impacts the electronic properties at temperatures above the magnetic transitions  producing anomalous temperature and doping dependences not present without alignment to hBN.  Close to the Mott transition we find that onsite interactions promote an antiferromagnetic (AF) state, probably breaking the C$_3$ symmetry, that will compete with the ferromagnetism arising from intersite exchange interactions to determine the ground state.

\end{abstract}
\maketitle
\section*{INTRODUCTION}
Since the discovery of insulating phases and superconductivity in magic angle twisted bilayer graphene (MATBG)~\cite{CaoNat2018_1,CaoNat2018_2}, moir\'e structures with narrow bands have become new platforms to explore electronic  correlations.  Insulating and, in some cases, superconducting states have been found in other moir\'e systems ~\cite{ChenNatPhys2019,LiuNat2020,TangNature2020,ParkNat2021,HaoScience2021,zhang2021ascendance,park2021magicangle}.  A key issue is whether the local correlations, which produce a plethora of unconventional behaviors in correlated materials such as high-Tc superconductors or heavy fermion materials
\cite{KeimerNature2015,ProustAnnRev2019,BasconesCRP2016,ChubukovPhysicsToday2015,Coleman-chapter,SiScience2010}, in some cases including Mott insulating states, are of relevance in moir\'e systems or their correlated states can be well explained in terms of standard band instabilities.

\begin{figure*}
\includegraphics[clip,width=\textwidth]{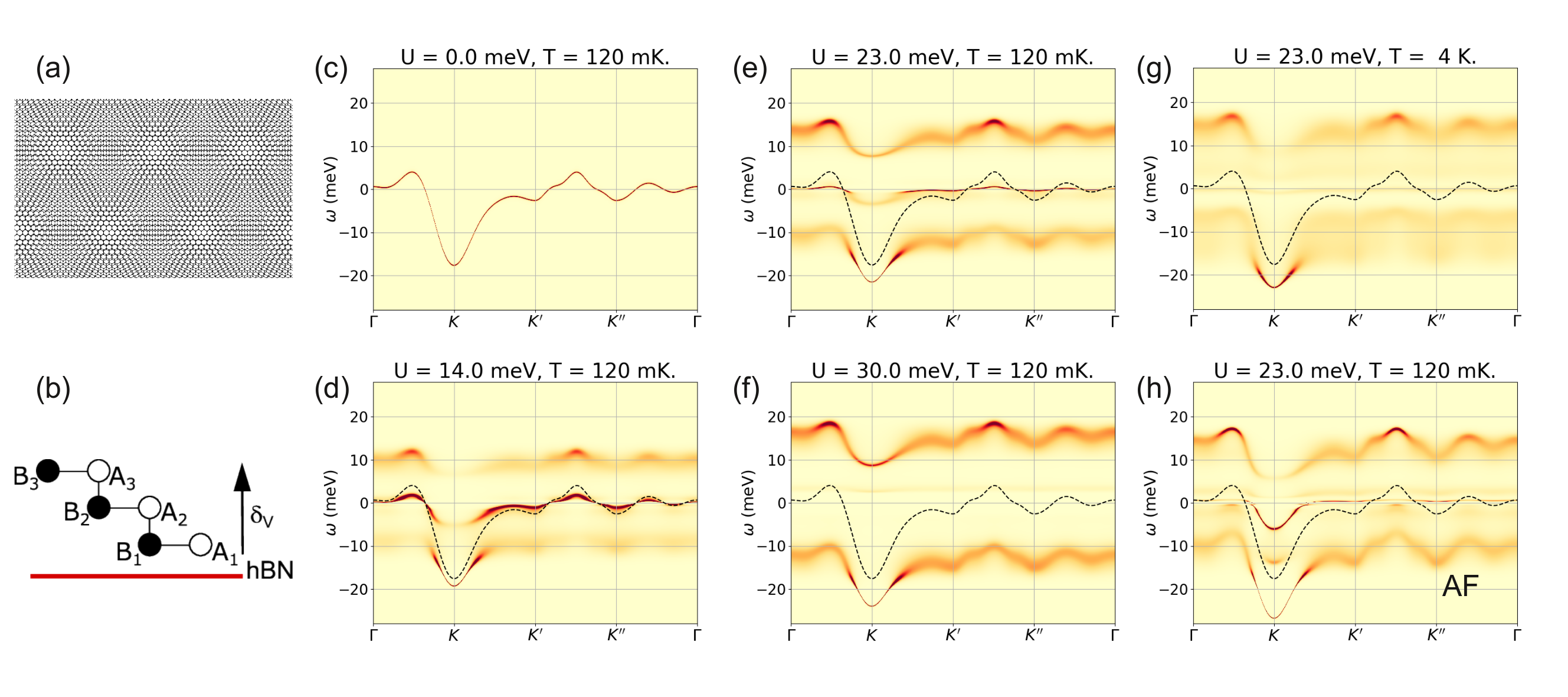}
\caption{{\bf System description, evolution of the band structure and distribution of the electronic spectral weight.}  (a) Moir\'e pattern created by a lattice mismatch between two honeycomb layers.  Not to scale. The moir\'e scale in ABC on hBN is $a_M \approx $15 nm, inversely proportional to the relative difference in lattice constants. 
(b) Sketch of the ABC trilayer stacking on hBN. A$_i$ and B$_i$ refer to the carbon sublattices in each graphene layer. Only one of the layers is aligned to hBN. 
(c) to (h) Spectral weight on the valence band of a given valley for $\delta_V=30$ meV at half-filling $n=-2$ . Each curve is centered at its corresponding chemical potential $\mu$. The dashed line in (d) to (h) corresponds to the 
non-interacting band in (c) and it is given as a guide to the eye. (c) to (f) correspond to U$=0, 14, 23$ and $30$ meV at T=$120$ mK. The symmetry breaking tendencies have been suppressed in the DMFT calculation by imposing 
an SU(4) symmetric solution.   As U increases, the quasiparticle band gets renormalized, the Fermi velocity is reduced and the electronic states have finite probability to decay away from the chemical potential $\mu$, giving a 
blurred aspect to the band, and part of the spectral weight is transferred to the incoherent Hubbard bands (around $\pm \rm{U}/2$). The quasiparticle band disappears at the Mott transition at U$_{{\rm Mott}}\sim 28$ meV. (g) Bands 
with U$= 23$ meV at T$= 4$ K. (h) Bands in the broken symmetry state assuming a spin antiferromagnet for U$=23$ meV at T$=120$ mK.} 
\label{fig:Fig1} 
\end{figure*}

ABC/hBN, the first moir\'e system to show insulating states shortly after MATBG~\cite{ChenNatPhys2019}, is playing a prominent role in this study. The moir\'e potential  
originates in the alignment between graphene and hBN, both with honeycomb structures but slightly different lattice constants, Figs.~\ref{fig:Fig1} (a)-(b). Comparing the experimental signatures in the presence and absence of such alignment facilitates isolating the effect of the moir\'e confinement  on  the correlated states. 
ABC trilayer graphene features bands with flat sections close to the charge neutrality point (CNP), owing to their cubic dispersion in the absence of trigonal warping~\cite{GuineaPRB2006,KoshinoPRB2009,ZhangPRB2010}.  A perpendicular electric field induces  a potential drop   $\delta_V$  between the top and the bottom graphene layers opening a gap at the CNP and making the bands even flatter. 
If the ABC trilayer graphene is aligned to hBN, 
the superlattice potential  gaps the bands at the boundaries of the moir\'e mini-Brillouin zone giving rise to narrow bands, isolated in the presence of a perpendicular electric field~\cite{ChenNatPhys2019}.  The bandwidth, the filling and the topological properties of these bands can be further tuned by gates, being the valence band narrower than the conduction band.  Depending on the electric field  orientation with respect to the hBN, the valence band can be topologically trivial (Chern number $=0$) or not (Chen number$= \pm 3$)~\cite{ZhangPRB2019b,ChittariPRL2019}.  Correlated insulating states have been observed in aligned samples for both signs of the electric field and one or two holes added to the valence band~\cite{ChenNatPhys2019,ChenNat2020,ZhouNat2021}. Due to the valley and spin degeneracy, up to four holes can be accomodated and 
half-filling corresponds to $n=-2$. Evidence of correlations such as ferromagnetism has also been observed at some metallic fillings~\cite{Chen2020,ZhouNat2021}.

Recent compressibility and Shubnikov de Haas measurements  showed a variety of spin and/or valley polarized metallic phases and even superconducting regions in non-aligned ABC graphene trilayers under a vertical displacement field for electronic densities similar to those of the correlated states in ABC/hBN~\cite{ZhouNat2021, ZhouNat2021b}.   In non-aligned ABC trilayer there is no moir\'e confining potential, ruling out any role of Mott physics at such low densities. The ordered phases were ascribed to the large density of states (DOS) enhanced by the electric field, with ferromagnetism originating in a Stoner instability~\cite{ZhouNat2021, ZhouNat2021b}.  Compressibility measurements in ABC/hBN exhibit strong similarities to the ones in non-aligned samples~\cite{ZhouNat2021}. Important differences induced by the alignment include the emergence of incompressible states at integer fillings, e.g. one and two holes/electrons per moir\'e unit cell, not allowed without the moir\'e, and the lack of symmetry with respect to the displacement field sign. Relying on the similarities, Ref.~\cite{ZhouNat2021} concluded that the correlated states in ABC/hBN, including the insulating ones, originate in an itinerant ferromagnetic Stoner instability,  with the moir\'e potential and band isolation being just a perturbation and Mott correlations playing essentially no role.  However, the specific features able to clarify the role of Mott physics~\cite{Fazekas-book1998,ImadaRMP1998,BasovRevModPhys2011} in ABC/hBN have not been addressed.

In this work we analyze the effect of the local correlations on the isolated topologically trivial valence band in ABC/hBN with the focus on the reorganization of the electronic spectral weight. 
The reshuffling of the spectral weight, which eventually leads to the Mott transition at integer fillings and high interactions,  is already prominent in the metallic state at interactions with
magnitudes that may be relevant in the experiments.  The strong band narrowing and the incoherent spectral weight  produced by the local correlations results in unconventional doping dependent 
electronic properties and may alter the ordering instabilities of the system.  
We have found that close to the Mott transition the intra moir\'e unit-cell interactions  promote an antiferromagnetic ordering in ABC/hBN
that probably breaks $C_3$ symmetry and that will compete with the ferromagnetic intersite exchange interactions to determine the ground state. The reorganization of the spectral weight 
induced by Mott correlations does not require symmetry breaking and the incoherence is enhanced with increasing temperature. We propose to study the doping and temperature dependent electronic properties above the ordering transitions as the best way to clarify the role of Mott physics  in ABC/hBN.

\section*{Model}
The topologically trivial character of the band allows for a description in terms of exponentially localised Wannier orbitals for each valley separately, facilitating the use of techniques employed in other strongly correlated systems. Assuming that the aligned hBN is below the ABC  trilayer graphene as in Fig.~\ref{fig:Fig1}(b), the valence band is trivial for $\delta_V>0$. This is equivalent to having $ \delta_V<0$ with hBN on top of the ABC trilayer~\cite{ZhangPRB2019b,ChittariPRL2019}. Following Ref.~\cite{ZhangPRB2019b} we model the trivial band as a two orbital system centered on an effective triangular moir\'e lattice, each of the orbitals corresponding to a given valley. The two-valleys are related by time reversal symmetry and are assumed to be decoupled. Within a valley the kinetic term includes hoppings up to four nearest neighbours, with hopping parameters depending on $\delta_V$, see Supplemental Material (SM) and Ref.~\cite{ZhangPRB2019b}.
Except otherwise stated, we assume a potential drop through the trilayer $\delta_V=30$ meV.  
According to the estimates of Ref.~\cite{ZhangPRB2019b}, for this value of $\delta_V$ the gaps between the trivial narrow valence band and other bands are   larger than $20$ 
 meV and assuming an isolated band is a reasonable approximation. The non-interacting electronic band and the density of states (DOS), plotted in 
 Fig.~\ref{fig:Fig1}(c) and Fig.~\ref{fig:Fig2}(a) respectively, feature several van Hove densities at fillings $n \sim -0.86, -0.96, -1.62$ and $-2.84$, relatively far from the range of densities around half-filling $n \sim -2$, which is the focus of this work.

The onsite U$_0=25$ meV and the first-nearest neighbor U$_1=10$ meV  density-density interactions estimated in Ref.~\cite{ZhangPRB2019b}  using a relative dielectric constant $\epsilon=8$ are much larger than the exchange and assisted hopping parameters $\leq 0.50$ meV.  Such a difference justifies keeping only the density-density terms on a first approximation. In this approximation the interactions are SU(4) symmetric. We will go back to the role of the smaller exchange terms in the discussion section. The onsite density-density interaction U$_0$ is considerably smaller than the one expected in MATBG U$_0^{\rm{MATBG}} \sim 60$ meV if the same $\epsilon=8$ is used~\cite{CalderonPRB2020}, suggesting a smaller tendency towards localization  in ABC/hBN than in MATBG.
Nevertheless, U$_0$ is  comparable to the bandwidth W, see Fig.~\ref{fig:Fig2}(a), and Mott correlations could be relevant. In the absence of charge modulation, the effect of the intersite density-density interaction U$_1$ on the Mott transition can be approximately accounted for by a reduction of the onsite interaction~\cite{SchulerPRL2013,HuangPRB2014}. In the following we model the ABC/hBN using a Hubbard model with an onsite effective interaction U and study the effect of correlations by performing single-site DMFT calculations~\cite{GeorgesRMP1996,kotliarRMP2006} using a continuous time Quantum Monte Carlo impurity solver~\cite{GullRMP2011, HaulePRB2007}. DMFT provides us with a suitable approach to study Mott physics, including the metal-insulator transition at integer fillings,  and permits analyzing the effect of correlations with and without symmetry breaking. While many Mott insulators order in one way or another at low temperatures, the Mott transition does not require any symmetry breaking and the Mott character of the transition is more easily identified in the non-ordered state. 

\section*{RESULTS} 
\subsection{Non-ordered state}
\begin{figure*}
\includegraphics[clip,width=\textwidth]{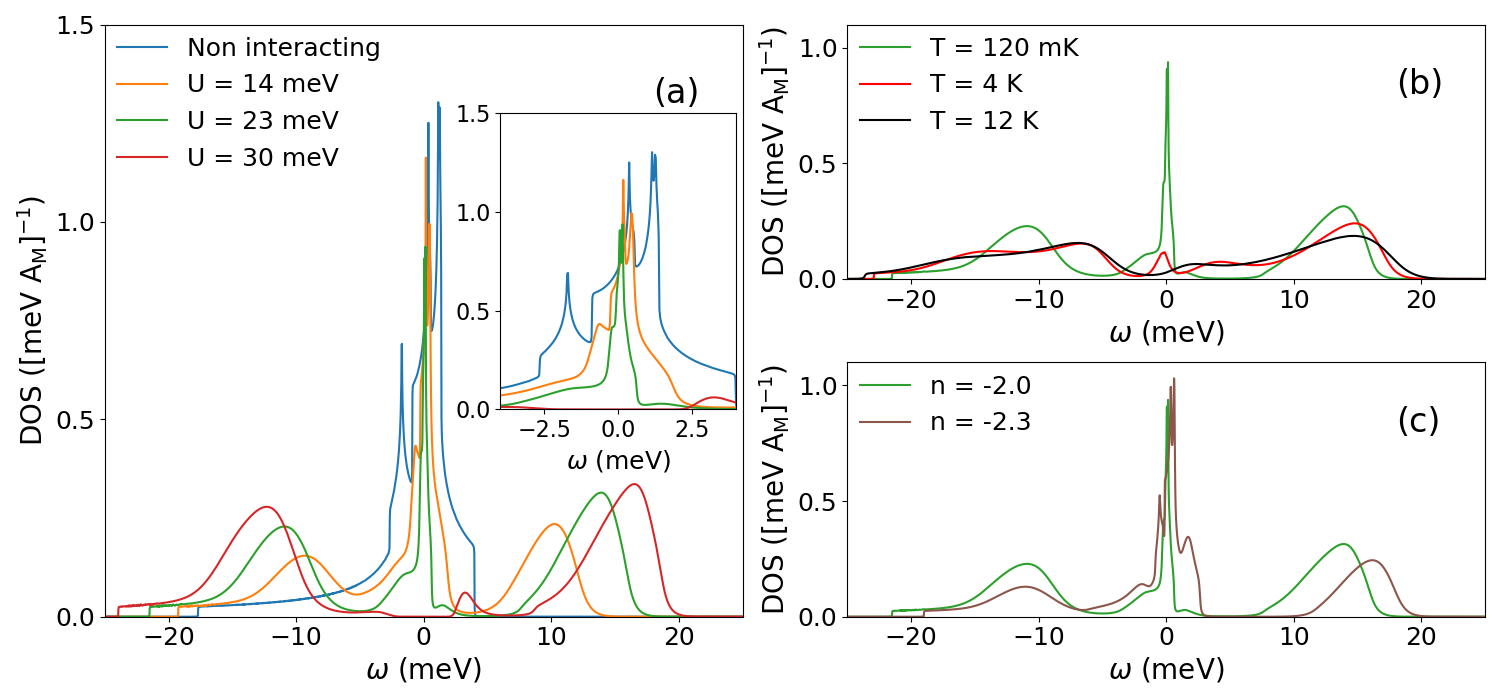}
\caption{{\bf Effect of interaction, temperature and filling on the density of states DOS}. Density of states (a) for different values of U at T=120mK at half-filling $n=-2$. Inset: zoom at low frequencies. The non-interacting case, in blue, is shown as a reference; (b) for different temperatures at U$=23$ and $n=-2$; (c) for different fillings at U$=23$ meV and T=120 mK. Each curve is centered at its corresponding chemical potential.   }
\label{fig:Fig2} 
\end{figure*}
Figs.~\ref{fig:Fig1} (c)-(f) show the spectral weight $A(\bf{k},\omega)$ for one of the valleys as a function of interaction U at half-filling $n=-2$ and low temperature (T$=120$ mK) in the non-ordered state. The symmetry breaking tendencies have been suppressed in the DMFT calculation by imposing an SU(4) symmetric solution. The energy of the states in the non-interacting band in Fig.~\ref{fig:Fig1}(c) are well defined, the band is sharp and thin, as the single-electron states are eigenstates of the Hamiltonian and do not decay. At U$=14$ meV, smaller than the bandwidth W$\sim 20$ meV, the effect of the electronic correlations on the band shape is already quite evident, Fig.~\ref{fig:Fig1}(d). The single electron states are no longer eigenstates of the interacting Hamiltonian and its lifetime is now finite. Two features stand out: the appearance of two incoherent bands separated by an energy gap $\sim$U and a strongly renormalized band at the Fermi energy. The incoherent bands, known as Hubbard bands, are not well defined. The renormalized band around the Fermi level is better defined, its effective mass has been enhanced, and its width is finite but small. These are heavy quasiparticle states, a truly many-body collective superposition of single particle states.
The effect of the correlations is more evident at U$=23$ meV in  Fig.~\ref{fig:Fig1}(e). At this value, the quasiparticle band around the chemical potential 
$\omega=0$ has become very narrow and has lost most of its spectral weight, while the Hubbard bands are well formed. 
At U$=30$ meV,  in the Mott insulating side, there is no band crossing the chemical potential, Fig.~\ref{fig:Fig1}(f). 
The quasiparticle band disappears at the Mott transition at U$_{{\rm Mott}} \sim 28$ meV. 

The strong reorganization of the spectral weight is manifest in the density of states which shows the characteristic three peak 
structure of the correlated metals approaching the Mott transition for U$<$U$_{{\rm Mott}}$~\cite{GeorgesRMP1996,kotliarRMP2006}, see Fig.~\ref{fig:Fig2}(a). In accordance with the effect 
observed in the bands, the quasiparticle peak around the chemical potential narrows with increasing interactions losing spectral weight  
which accumulates in the two bumps at positive and negative energies corresponding to the Hubbard bands. In the non-interacting case, around  the 81$\%$ of the total spectral weight  is  within $-3$ and $4$ meV from the 
chemical potential, but this percentage is reduced to 41$\%$ and 19$\%$ at U$=14$ and $23$ meV, respectively. At  U$=30$ meV, above U$_{{\rm Mott}}$, the quasiparticle peak at 
the chemical potential is absent. Almost all the spectral weight has been transferred to the Hubbard bands.

The incoherence grows with increasing temperature. This is already evident at a few Kelvin. 
At $4$ K, a temperature much smaller than the non-interacting bandwidth,  the quasiparticle band  
at U$=23$ meV cannot be distinguished in Fig.~\ref{fig:Fig1}(g). Concomitantly, the peak in the 
DOS at the chemical potential is suppressed, Fig.~\ref{fig:Fig2}(b).  
Doping away from integer filling favors the metallic states and
correlations induce changes in
the DOS and a non-rigid bandshift.
 In Fig.~\ref{fig:Fig2}(c), 
the width of the quasiparticle peak in the DOS
is larger at  $n=-2.3$  than at $n=-2$, showing closer resemblance, also in shape, to the non-interacting DOS in  Fig.~\ref{fig:Fig2}(a). At U$=14$ meV, considerably 
far from the Mott transition, the enhancement of the incoherence at a few kelvin and the change of the quasiparticle peak width with doping are less pronounced but still visible, see SM.

\subsection{Ordered state}
\begin{figure*}
\includegraphics[clip,width=\textwidth]{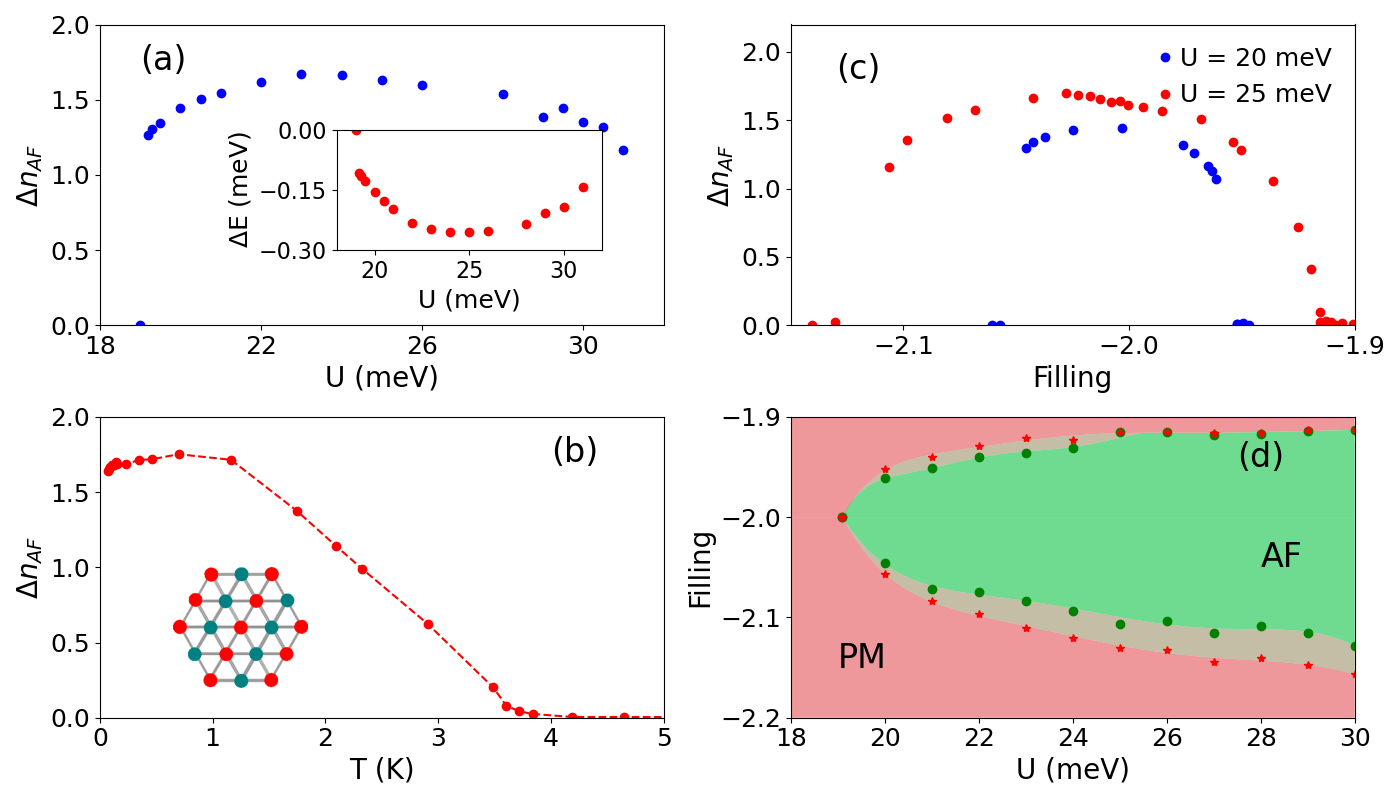}
\caption{{\bf Broken symmetry states.}  (a) Order parameter $\Delta n_{\rm AF}$ in the AF state as a function of U  at half-filling and T$=120$ mK.  The inset shows the corresponding energy gain in  the broken symmetry state with respect to the non-ordered one.  (b) $\Delta n_{\rm AF}$  at half-filling as a function of temperature for U$=23$ meV. (c)  $\Delta n_{\rm AF}$  as a function of the filling around  $n=-2$ for U$=20$meV and U$=25$meV.   (d) Phase diagram as a function of the filling $n$ and the interaction U. In the AF green region, $\Delta n_{\rm AF} \ne 0$, and in the red PM one, $\Delta n_{\rm AF} = 0$. The shaded area between the two regions is inhomogeneous, see SM. }
\label{fig:Fig3} 
\end{figure*}

We now allow for the breaking of spin and/or valley symmetry. 
The Hubbard model has a strong tendency towards antiferromagnetism at half-filling at large enough interactions.
However, the triangular lattice is frustrated as it cannot satisfy that all nearest neighbors are antiferromagnetically ordered. At low temperatures and large U, the ground state of the single orbital Hubbard model on a triangular lattice with hopping restricted to first nearest neighbors is a coplanar $120^\circ$ antiferromagnet, but this order may be destabilized by the longer range hoppings or their complex phases and other states such as spin liquids or C$_3$ symmetry breaking AF order could be favoured ~\cite{WietekPRX2021,RohringerRMP2018}. Within a certain range of interactions a large density of states could also stabilize a 
ferromagnetic state. Starting from a Wannier function model for ABC/hBN analogous to the one used here but corresponding to $\delta_V=20$ meV and performing Variational Monte Carlo calculations (VMC), Chen et al~\cite{Chen2020a} found a C$_3$ symmetry breaking AF ground state at $n=-2$, as illustrated in the inset of Fig.~\ref{fig:Fig3}(b). 
In their calculations, the SU(4) symmetry of the interactions was broken by an intra-valley Hund's coupling term J$_{\rm{H}}$. Within the range of parameters studied by these authors, a finite value of J$_{\rm{H}}$  seems necessary to stabilize the AF state. The ground state found was a spin-antiferromagnet, while an antiferro-valley ground state was not stabilized~\cite{Chen2020a}. 

 In our calculations, at $n=-2$, we have not found ferromagnetism for any interaction U below $30$ meV but an  AF ordered state with broken SU(4) spin-valley symmetry.  
 J$_{\rm{H}}$ or any other SU(4) symmetry breaking interaction terms are taken equal to zero. Correspondingly,  the spin-antiferromagnet and the antiferro-valley ordering are degenerate and equally stable. 
Fig.~\ref{fig:Fig3}(a) shows the order parameter $\Delta n_{\rm AF}$ as a function of interaction. 
 $\Delta n_{\rm AF}$, with a maximum possible value of $2$, is defined as the intra-unit cell difference in occupation between the two valleys  $\Delta n_{\rm AF}= \sum_{\sigma} (n_{ \zeta,\sigma}-n_{\bar \zeta,\sigma})$ in the pure antiferrovalley state or the two spins $\Delta n_{\rm AF}= \sum_{\zeta} (n_{ \zeta,\sigma}-n_{\zeta,\bar \sigma})$ in the pure spin AF, but due to the SU(4) symmetry other combinations of spin $\sigma$ and valley $\zeta$ antiferromagnetism are possible. 
 The ordering at $n=-2$ emerges at an interaction $\sim 30\%$ smaller than  U$_{{\rm Mott}}$ and remains in the Mott insulator up to at least $31$ meV. 
 We found difficulties to converge the AF state beyond this interaction and we cannot confirm whether this state is stable at larger U.  The magnetic order persists up to temperatures $\sim 3-4$~K, Fig.~\ref{fig:Fig3}(b), compatible with the energy gain due to the antiferromagnetism  $\sim 0.25$meV, see inset of Fig.~\ref{fig:Fig3}(a).  On spite of the small energy gain, much smaller than the effective interaction U, the AF ordering induces changes in the spectral weight in a large energy window, compare Fig.~\ref{fig:Fig1}(e) and Fig.~\ref{fig:Fig1}(h).   We have found that the AF state is weakly metallic below U$_{{\rm Mott}}$, but it is not a Fermi liquid very close to the transition, see SM.  
Above U$\sim 20.5$ meV the kinetic energy decreases  in the AF state while the potential energy increases. This suggests that the AF is stabilized by super-exchange processes, in accordance with the AF being stable only close to the Mott insulating state and the doping dependence. Doping away from half-filling quickly suppresses the ordering, see Fig.~\ref{fig:Fig3}(c) and (d), with antiferromagnetism surviving to changes in the filling of only  $\sim \pm 0.1$ holes per moir\'e unit cell.

 We have found an AF state, with the same characteristics to the one 
 discussed above, starting from the bands of ABC/hBN corresponding to $\delta_V=20$ meV and  $\delta_V=50$ meV, not shown. On the other hand, the order is absent in the two-orbital Hubbard model on the triangular lattice with real hopping only to nearest neighbors. This suggests an important role of the complex long range hoppings in the 
 antiferromagnetism.  Within the DMFT calculations performed here, we cannot identify the pattern characterizing the AF ordering. Nevertheless, considering the states that can be stabilized within the single site DMFT calculations performed, which do not include the $120^{\circ}$ antiferromagnet, we 
 believe that the AF ordering found has the C$_3$ symmetry breaking pattern found by Chen et al~\cite{Chen2020a}, Fig.~\ref{fig:Fig3}(c).  
The lack of magnetic order at  J$_{\rm{H}}=0$ in~Ref.\cite{Chen2020a} is most probably due to the specific value of the interaction $\rm U=0.93\tilde W$  studied in that work, with $\rm \tilde W$ the $\delta_V=20$meV bandwidth. This interaction is considerably smaller than the interaction  $\rm \tilde U_{{ Mott}} \sim1.5$$\rm \tilde W$ at which, according to our calculations, the Mott transition takes place for $\delta_V=20$meV and J$_{\rm{H}}$=0.  We have not found AF at n$=-2$ for $\rm U=0.93\tilde W$ 
and $\delta_V=20$ meV in agreement with Ref.~\cite{Chen2020a}.

\section*{DISCUSSION}

The insulating states at integer filling in ABC/hBN~\cite{ChenNatPhys2019,ChenNat2020,ZhouNat2021} were proposed to originate in a Stoner instability~\cite{ZhouNat2021}.
Producing an insulating state through spin and/or valley polarization should involve interactions comparable or larger than the bandwidth.
Assuming an isolated flat valence band we have shown that for interactions of such magnitude local correlations are expected to induce a strong reorganization of the electronic spectral weight in ABC/hBN, more pronounced at  integer fillings of the moir\'e unit cell. The spectral weight reshuffling, which is one of the main signatures of Mott physics, happens not only at interactions able to drive the system to a Mott insulating state but also at intermediate values of U. In the latter case the system behaves as a correlated metal at temperatures above any symmetry breaking transition. If Mott correlations are relevant in ABC/hBN they will show up in the non-ordered state, insulating or metallic, through measurable unconventional doping and temperature dependent electronic properties. Moreover, the predictions of weak coupling approaches regarding the system tendency to a band instability  may be altered by the spectral weight loss of the quasiparticle band and the band shape modification induced by the local correlations. The spectral weight reorganization discussed here is only expected in the presence of the moir\'e confining potential. Therefore such unconventional behavior should not be present in the non-aligned ABC graphene trilayer. We leave for future work a more detailed theoretical study of these effects.

The spectral weight transfer is not expected to depend qualitatively on the details of the non-interacting band, but the latter could have some influence on the magnetic orderings.
In the calculations we neglected the onsite and intersite exchange terms. These terms were estimated to be much smaller than the density-density interactions and therefore they are not expected to alter significantly the non-ordered state~\cite{ZhangPRB2019b}. On the other hand they are comparable to the energy gain due to the antiferromagnetic ordering found and could modify the predictions in the ordered state.  On general grounds a finite value of the Hund's coupling interaction J$_{\rm H}$ breaks SU(4) symmetry such that spin and valley ordering would be non-degenerate. Whether spin or valley is promoted depends on the sign of J$_{\rm H}$.  Ref.~\cite{ZhangPRB2019b} estimated  J$_{\rm H}$ $\approx$0.1 meV and positive but its value is currently under discussion~\cite{ZhouNat2021}. 

The small overlap between the wave functions at neighboring unit cells produces an intersite exchange interaction I$_{\rm H}\approx 0.4$ meV~\cite{ZhangPRB2019b}. This interaction favors parallel alignment of the spins and would compete with the antiferromagnetic ordering due to superexchange found here. Ferromagnetic order was found in the VMC calculations of Chen et al~\cite{Chen2020a} for $\delta_V$=20 meV at values of I$_{\rm H}$ larger than $0.3$ meV. Ferromagnetism is therefore a plausible ground state even in proximity to the Mott transition or in the Mott  insulating state. On the other hand, we note that the energy difference between the ferromagnetic and antiferromagnetic orderings could be comparable to the energy scales of external perturbations such as magnetic fields or strains, opening the possibility to tune a transition between different magnetic states by external knobs. Finding experimentally such a transition would also shed light on the nature of correlations in ABC/hBN. In summary, more experiments are needed to conclude whether Mott correlations are relevant in ABC/hBN. 

{\it Note added upon submission:} On the same day on which we are submitting the manuscript, a new experimental preprint studying the photocurrent spectrum of ABC/hBN has appeared in the arXiv (Science in press)~\cite{yang2022}. In this work, a broad absorption peak, absent at charge neutrality, is observed at half-filling of the valence band at $\sim18$ meV,  indicating a direct optical excitation across an emerging Mott gap. This experimental observation supports our picture. Further predictions for the photocurrent from the evolution of the DOS with doping in our work (see Fig.~\ref{fig:Fig2}(c)) include (i) the expected presence of the correlated absorption peak, but with reduced intensity with respect to the one observed at half-filling, at metallic dopings close to half-filling; and (ii) the narrowing of the quasiparticle peak as half-filling is approached should be observable in the doping dependent width of the absorption peak between the moir\'e bands  in the metallic correlated state.

  \renewcommand{\thesection}{SUPPLEMENTAL MATERIAL}%
        \setcounter{table}{0} 
        \renewcommand{\thetable}{S\arabic{table}}%
        \setcounter{figure}{0}
        \renewcommand{\thefigure}{S\arabic{figure}}%
          \setcounter{equation}{0}
        \renewcommand{\theequation}{S\arabic{equation}}%
        
\section*{Supplemental material}

We use the prescription of Ref.~\cite{ZhangPRB2019b} which starts from the continuum model of a graphene ABC trilayer aligned to an hBN layer and projects the trivial valence band onto a triangular lattice Wannier function model.  
For  $\delta_V=30$ meV, the  hopping parameters corresponding to a given valley are   $t_1= -1.227 e^{i 0.249\pi}$, $t_2=0.879$, $t_3=-0.267 e^{i0.1\pi}$, and $t_4=-0.61 e^{-i 0.599 \pi}$, all in meV. The parameters in the other valley can be obtained by time reversal symmetry. The sign of the hoppings here is opposite to the one in Ref.~\cite{ZhangPRB2019b} as they work in the hole picture.  To facilitate comparison with experiment, we have made explicit that the isolated band under consideration is a valence band. The values of the interaction were obtained in Ref.~\cite{ZhangPRB2019b} projecting the screened Coulomb interaction on to the valence band, assuming a dielectric constant $\epsilon=8$ and a screening length $\sim 75 $ nm.  More details on the model, including the hopping parameters corresponding to other values of $\delta_V$, can be found in Ref.~\cite{ZhangPRB2019b}.

\begin{figure*}
\includegraphics[clip,width=0.92\textwidth]{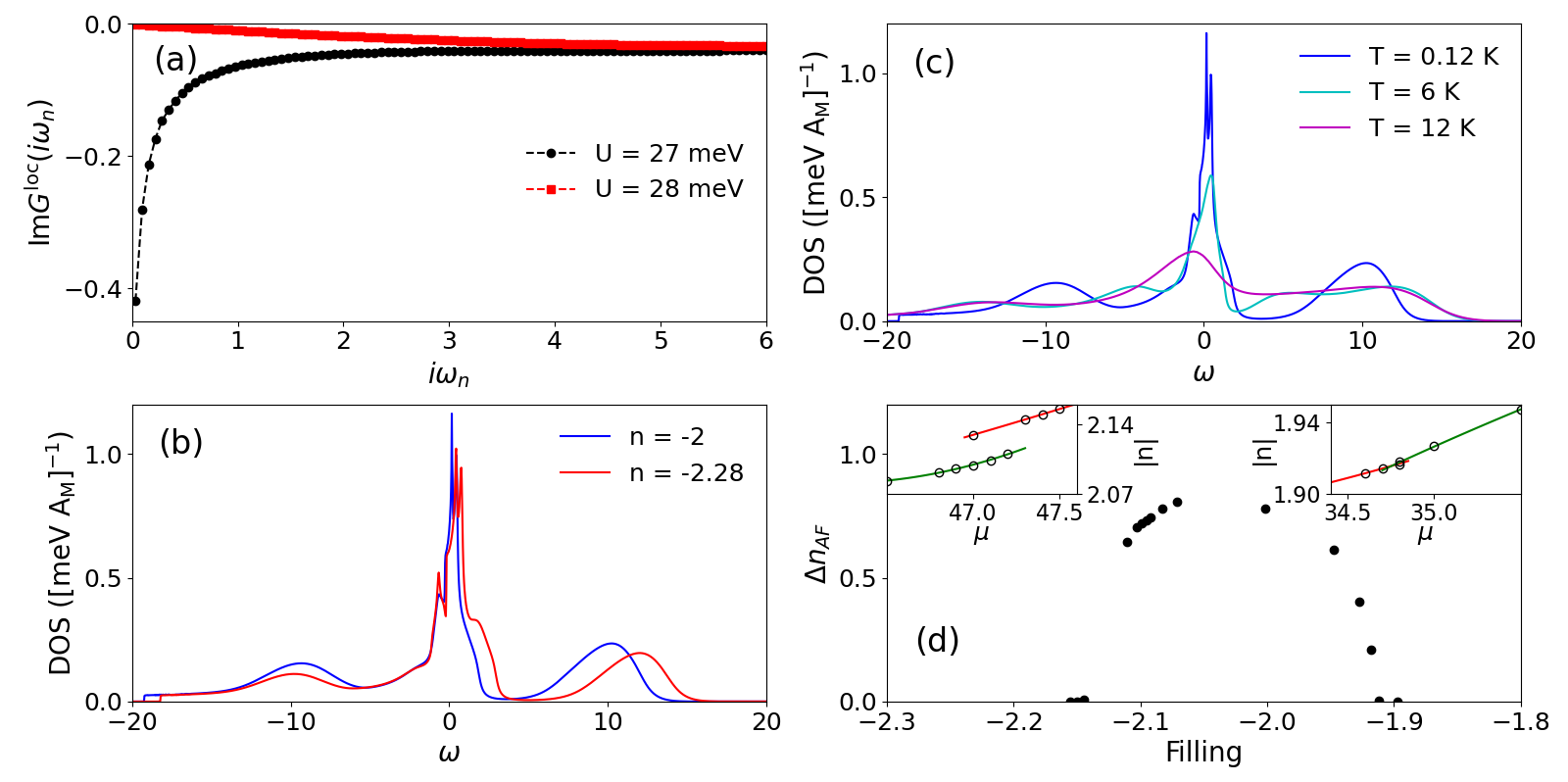}
\caption{ a) Imaginary part of the local Green's funcion in Matsubara frequency $G^{loc}(i\omega_n)$ at $n=-2$  and T=120 mK  in the non magnetic state evidencing the Mott  metal-insulating transition between U=27 and U=28 meV.   (b) DOS for U$=14$ meV at T$=120$ mK for different fillings. (c) DOS for U$=14$ meV at $n=-2$ for different temperatures. (d) $\Delta n_{AF}$ versus filling for U$=27$ meV. Doping with holes from $n=-2$ leads to a region where two solutions with different densities exist at the same $\mu$: an AF solution, with finite $\Delta n_{AF}$, and an PM one, with  $\Delta n_{AF}=0$ (see inset on the left, with the red curve corresponding to the PM solution and the green one to the AF one). At the transition there is a jump in the order parameter. Doping with electrons the density and the order parameter evolve continuously at the transition, see inset on the right.}
\label{fig:FigS} 
\end{figure*}

We perform the single-site DMFT calculations  at a given chemical potential $\mu$  using a continuous time quantum Monte Carlo impurity solver as implemented in Ref.~\cite{HaulePRB2007}. As output, we obtain the self-energy and the impurity Green's function in Matsubara frequency $\Sigma_{\zeta \sigma}(i \omega_n)$ and $G^{\mathrm{imp}}_{\zeta \sigma}(i \omega_n)$, with $\zeta$ and $\sigma$ respectively referring to the valley and the spin degrees of freedom. 
From the self-energy, the lattice Green's function can be calculated as 
\begin{equation}
G_{\zeta \sigma}(\mathbf{k}, i \omega_n) =\frac{1}{i\omega_n -(\epsilon_{\zeta \bf k}-\mu)- \Sigma_{\zeta \sigma}(i \omega_n) }.
\end{equation}

When convergence is reached, the self-consistency condition imposes $G^{\mathrm{imp}}_{\zeta \sigma} = G^{\mathrm{loc}}_{\zeta \sigma}$ with $G^\mathrm{loc}_{\zeta \sigma}(i \omega_n) = \sum_\mathrm{k}G_{\zeta \sigma}(\mathbf{k}, i \omega_n)$ \cite{GeorgesRMP1996}. 
  In single-site DMFT, $\Sigma_{\zeta \sigma}(i \omega_n)$  depends on frequency but not on momentum.  
In the absence of symmetry breaking, $\Sigma_{\zeta \sigma}(i\omega_n)$ and $G^{loc}_{\zeta \sigma}(i\omega_n)$ do not depend on the spin or the valley and we can drop the corresponding indices.

We determine the metallic or insulating character in both the non-ordered and the ordered states from the imaginary part of the local Green's function in Matsubara frequencies; Im$G^{loc}_{\zeta \sigma}(i \omega_n)$ tends to zero at low frequencies in the insulator and to a finite value in the metal. Fig.~\ref{fig:FigS}(a) shows Im$G^{loc}_{\zeta \sigma}(i \omega_n)$  at U$=27$ meV and U$=28$ meV evidencing the Mott metal-insulator transition. In the non-ordered metallic state, Im$\Sigma(i \omega_n)$ vanishes linearly with $\omega_n$ as $i\omega_n \rightarrow 0$ as expected in a Fermi liquid at very low temperatures. On the other hand, in the metallic AF state,  only for interactions below U$=23$ meV, Im$\Sigma_{\zeta \sigma}(i \omega_n)$ approaches linearly zero at $i\omega_n \rightarrow 0$  evidencing Fermi liquid behavior. Above this value of U, Im$\Sigma_{\zeta \sigma}(i \omega_n)$  tends to a finite value as  $i\omega_n \rightarrow 0$, not shown. This behavior implies that above U$=23$ meV the quasiparticle states have a finite lifetime at the chemical potential, i.e. non Fermi-liquid behavior.  

We perform the analytic continuation of  $\Sigma_{\zeta \sigma}(i \omega_n)$ to obtain the real frequency self energy $\Sigma_{\zeta \sigma}(\omega)$ using the maximum entropy method \cite{JarrelPR1996}. The spectral weight $A_{\zeta \sigma}({\bf k},\omega)$ and the density of states $\rho (\omega)$ discussed in the main text are given by $A_{\zeta \sigma}({\bf k},\omega)={-\frac{1}{\pi}}{\rm Im} G_{\zeta \sigma}({\bf k},\omega)$ and $ \rho (\omega)=\sum_{\zeta \sigma {\bf k}} A_{\zeta \sigma}({\bf k},\omega)$.

Figs.~\ref{fig:FigS}(b) and (c) show the density of states at U$=14$ meV for different temperatures and dopings, respectively. With increasing temperature, the suppression of the quasiparticle peak is less pronounced than at U$=23$ meV in Fig.~2(b) in the main text.  As in  Fig.~2(c), doping away from half-filling reduces the narrowing of the quasiparticle peak in the density of states. The effect is less noticeable than at similar doping levels at U$=23$ meV. 

Antiferromagnetism emerges as a systematic alternation in the DMFT iterations of the spin and/or valley dependent self-energy, producing different occupations in the different degrees of freedom and a finite $\Delta n_{\rm AF}$. The phase diagram in Fig. 3(d) has been calculated varying the chemical potential $\mu$. In general, for a fixed $\mu$  the density  in the AF and the non-magnetic PM states is different and a jump in the order parameter 
$\Delta n_{\rm AF}$ is accompanied by a jump in the density. Fig.~\ref{fig:FigS}(d) illustrates, for U$=27$ meV, two typical behaviours found when plotting the order parameter as a function of filling: in this case, for hole doping there are AF and PM solutions for a given $\mu$ and a jump in $\Delta n_{\rm AF}$, while electron doping shows a continuous behaviour.  In the region of parameters where the density does not evolve continuously between both solutions, the  boundaries of the AF and PM regions in the phase diagram in Fig. 3(d) correspond to the densities at the chemical potential  $\mu$ at which the order parameter jumps. 
In the shaded areas between these two boundaries  phase separation, inhomogeneities and hysteretic behavior may be found.

\section*{DATA AVAILABILITY}
All relevant data are available from the authors upon reasonable request.

\section*{ACKNOWLEDGMENTS}
M.J.C and E.B. acknowledge funding from PGC2018-097018-B-I00 (MCIN/AEI/FEDER, EU).
\bibliography{trilayerfinal}

\begin{thebibliography}{39}%
\makeatletter
\providecommand \@ifxundefined [1]{%
 \@ifx{#1\undefined}
}%
\providecommand \@ifnum [1]{%
 \ifnum #1\expandafter \@firstoftwo
 \else \expandafter \@secondoftwo
 \fi
}%
\providecommand \@ifx [1]{%
 \ifx #1\expandafter \@firstoftwo
 \else \expandafter \@secondoftwo
 \fi
}%
\providecommand \natexlab [1]{#1}%
\providecommand \enquote  [1]{``#1''}%
\providecommand \bibnamefont  [1]{#1}%
\providecommand \bibfnamefont [1]{#1}%
\providecommand \citenamefont [1]{#1}%
\providecommand \href@noop [0]{\@secondoftwo}%
\providecommand \href [0]{\begingroup \@sanitize@url \@href}%
\providecommand \@href[1]{\@@startlink{#1}\@@href}%
\providecommand \@@href[1]{\endgroup#1\@@endlink}%
\providecommand \@sanitize@url [0]{\catcode `\\12\catcode `\$12\catcode
  `\&12\catcode `\#12\catcode `\^12\catcode `\_12\catcode `\%12\relax}%
\providecommand \@@startlink[1]{}%
\providecommand \@@endlink[0]{}%
\providecommand \url  [0]{\begingroup\@sanitize@url \@url }%
\providecommand \@url [1]{\endgroup\@href {#1}{\urlprefix }}%
\providecommand \urlprefix  [0]{URL }%
\providecommand \Eprint [0]{\href }%
\providecommand \doibase [0]{https://doi.org/}%
\providecommand \selectlanguage [0]{\@gobble}%
\providecommand \bibinfo  [0]{\@secondoftwo}%
\providecommand \bibfield  [0]{\@secondoftwo}%
\providecommand \translation [1]{[#1]}%
\providecommand \BibitemOpen [0]{}%
\providecommand \bibitemStop [0]{}%
\providecommand \bibitemNoStop [0]{.\EOS\space}%
\providecommand \EOS [0]{\spacefactor3000\relax}%
\providecommand \BibitemShut  [1]{\csname bibitem#1\endcsname}%
\let\auto@bib@innerbib\@empty
\bibitem [{\citenamefont {Cao}\ \emph {et~al.}(2018{\natexlab{a}})\citenamefont
  {Cao}, \citenamefont {Fatemi}, \citenamefont {Demir}, \citenamefont {Fang},
  \citenamefont {Tomarken}, \citenamefont {Luo}, \citenamefont
  {Sanchez-Yamagishi}, \citenamefont {Watanabe}, \citenamefont {Taniguchi},
  \citenamefont {Kaxiras}, \citenamefont {Ashoori},\ and\ \citenamefont
  {Jarillo-Herrero}}]{CaoNat2018_1}%
  \BibitemOpen
  \bibfield  {author} {\bibinfo {author} {\bibfnamefont {Y.}~\bibnamefont
  {Cao}}, \bibinfo {author} {\bibfnamefont {V.}~\bibnamefont {Fatemi}},
  \bibinfo {author} {\bibfnamefont {A.}~\bibnamefont {Demir}}, \bibinfo
  {author} {\bibfnamefont {S.}~\bibnamefont {Fang}}, \bibinfo {author}
  {\bibfnamefont {S.~L.}\ \bibnamefont {Tomarken}}, \bibinfo {author}
  {\bibfnamefont {J.~Y.}\ \bibnamefont {Luo}}, \bibinfo {author} {\bibfnamefont
  {J.~D.}\ \bibnamefont {Sanchez-Yamagishi}}, \bibinfo {author} {\bibfnamefont
  {K.}~\bibnamefont {Watanabe}}, \bibinfo {author} {\bibfnamefont
  {T.}~\bibnamefont {Taniguchi}}, \bibinfo {author} {\bibfnamefont
  {E.}~\bibnamefont {Kaxiras}}, \bibinfo {author} {\bibfnamefont {R.~C.}\
  \bibnamefont {Ashoori}},\ and\ \bibinfo {author} {\bibfnamefont
  {P.}~\bibnamefont {Jarillo-Herrero}},\ }\bibfield  {title} {\bibinfo {title}
  {Correlated insulator behaviour at half-filling in magic-angle graphene
  superlattices},\ }\href@noop {} {\bibfield  {journal} {\bibinfo  {journal}
  {Nature}\ }\textbf {\bibinfo {volume} {556}},\ \bibinfo {pages} {80}
  (\bibinfo {year} {2018}{\natexlab{a}})}\BibitemShut {NoStop}%
\bibitem [{\citenamefont {Cao}\ \emph {et~al.}(2018{\natexlab{b}})\citenamefont
  {Cao}, \citenamefont {Fatemi}, \citenamefont {Fang}, \citenamefont
  {Watanabe}, \citenamefont {Taniguchi}, \citenamefont {Kaxiras},\ and\
  \citenamefont {Jarillo-Herrero}}]{CaoNat2018_2}%
  \BibitemOpen
  \bibfield  {author} {\bibinfo {author} {\bibfnamefont {Y.}~\bibnamefont
  {Cao}}, \bibinfo {author} {\bibfnamefont {V.}~\bibnamefont {Fatemi}},
  \bibinfo {author} {\bibfnamefont {S.}~\bibnamefont {Fang}}, \bibinfo {author}
  {\bibfnamefont {K.}~\bibnamefont {Watanabe}}, \bibinfo {author}
  {\bibfnamefont {T.}~\bibnamefont {Taniguchi}}, \bibinfo {author}
  {\bibfnamefont {E.}~\bibnamefont {Kaxiras}},\ and\ \bibinfo {author}
  {\bibfnamefont {P.}~\bibnamefont {Jarillo-Herrero}},\ }\bibfield  {title}
  {\bibinfo {title} {Unconventional superconductivity in magic-angle graphene
  superlattices},\ }\href@noop {} {\bibfield  {journal} {\bibinfo  {journal}
  {Nature}\ }\textbf {\bibinfo {volume} {556}},\ \bibinfo {pages} {43}
  (\bibinfo {year} {2018}{\natexlab{b}})}\BibitemShut {NoStop}%
\bibitem [{\citenamefont {Chen}\ \emph {et~al.}(2019)\citenamefont {Chen},
  \citenamefont {Jiang}, \citenamefont {Wu}, \citenamefont {Lyu}, \citenamefont
  {Li}, \citenamefont {Chittari}, \citenamefont {Watanabe}, \citenamefont
  {Taniguchi}, \citenamefont {Shi}, \citenamefont {Jung}, \citenamefont
  {Zhang},\ and\ \citenamefont {Wang}}]{ChenNatPhys2019}%
  \BibitemOpen
  \bibfield  {author} {\bibinfo {author} {\bibfnamefont {G.}~\bibnamefont
  {Chen}}, \bibinfo {author} {\bibfnamefont {L.}~\bibnamefont {Jiang}},
  \bibinfo {author} {\bibfnamefont {S.}~\bibnamefont {Wu}}, \bibinfo {author}
  {\bibfnamefont {B.}~\bibnamefont {Lyu}}, \bibinfo {author} {\bibfnamefont
  {H.}~\bibnamefont {Li}}, \bibinfo {author} {\bibfnamefont {B.~L.}\
  \bibnamefont {Chittari}}, \bibinfo {author} {\bibfnamefont {K.}~\bibnamefont
  {Watanabe}}, \bibinfo {author} {\bibfnamefont {T.}~\bibnamefont {Taniguchi}},
  \bibinfo {author} {\bibfnamefont {Z.}~\bibnamefont {Shi}}, \bibinfo {author}
  {\bibfnamefont {J.}~\bibnamefont {Jung}}, \bibinfo {author} {\bibfnamefont
  {Y.}~\bibnamefont {Zhang}},\ and\ \bibinfo {author} {\bibfnamefont
  {F.}~\bibnamefont {Wang}},\ }\bibfield  {title} {\bibinfo {title} {{Evidence
  of a gate-tunable Mott insulator in a trilayer graphene moir{\'{e}}
  superlattice}},\ }\href {https://doi.org/10.1038/s41567-018-0387-2}
  {\bibfield  {journal} {\bibinfo  {journal} {Nature Physics}\ }\textbf
  {\bibinfo {volume} {15}},\ \bibinfo {pages} {237} (\bibinfo {year}
  {2019})}\BibitemShut {NoStop}%
\bibitem [{\citenamefont {Liu}\ \emph {et~al.}(2020)\citenamefont {Liu},
  \citenamefont {Hao}, \citenamefont {Khalaf}, \citenamefont {Lee},
  \citenamefont {Ronen}, \citenamefont {Yoo}, \citenamefont {Haei~Najafabadi},
  \citenamefont {Watanabe}, \citenamefont {Taniguchi}, \citenamefont
  {Vishwanath},\ and\ \citenamefont {Kim}}]{LiuNat2020}%
  \BibitemOpen
  \bibfield  {author} {\bibinfo {author} {\bibfnamefont {X.}~\bibnamefont
  {Liu}}, \bibinfo {author} {\bibfnamefont {Z.}~\bibnamefont {Hao}}, \bibinfo
  {author} {\bibfnamefont {E.}~\bibnamefont {Khalaf}}, \bibinfo {author}
  {\bibfnamefont {J.~Y.}\ \bibnamefont {Lee}}, \bibinfo {author} {\bibfnamefont
  {Y.}~\bibnamefont {Ronen}}, \bibinfo {author} {\bibfnamefont
  {H.}~\bibnamefont {Yoo}}, \bibinfo {author} {\bibfnamefont {D.}~\bibnamefont
  {Haei~Najafabadi}}, \bibinfo {author} {\bibfnamefont {K.}~\bibnamefont
  {Watanabe}}, \bibinfo {author} {\bibfnamefont {T.}~\bibnamefont {Taniguchi}},
  \bibinfo {author} {\bibfnamefont {A.}~\bibnamefont {Vishwanath}},\ and\
  \bibinfo {author} {\bibfnamefont {P.}~\bibnamefont {Kim}},\ }\bibfield
  {title} {\bibinfo {title} {Tunable spin-polarized correlated states in
  twisted double bilayer graphene},\ }\href
  {https://doi.org/10.1038/s41586-020-2458-7} {\bibfield  {journal} {\bibinfo
  {journal} {Nature}\ }\textbf {\bibinfo {volume} {583}},\ \bibinfo {pages}
  {221} (\bibinfo {year} {2020})}\BibitemShut {NoStop}%
\bibitem [{\citenamefont {Hao}\ \emph {et~al.}(2020)\citenamefont {Hao},
  \citenamefont {Zimmerman}, \citenamefont {Ledwith}, \citenamefont {Khalaf},
  \citenamefont {Najafabadi}, \citenamefont {Watanabe}, \citenamefont
  {Taniguchi}, \citenamefont {Vishwanath},\ and\ \citenamefont
  {Kim}}]{TangNature2020}%
  \BibitemOpen
  \bibfield  {author} {\bibinfo {author} {\bibfnamefont {Z.}~\bibnamefont
  {Hao}}, \bibinfo {author} {\bibfnamefont {A.}~\bibnamefont {Zimmerman}},
  \bibinfo {author} {\bibfnamefont {P.}~\bibnamefont {Ledwith}}, \bibinfo
  {author} {\bibfnamefont {E.}~\bibnamefont {Khalaf}}, \bibinfo {author}
  {\bibfnamefont {D.~H.}\ \bibnamefont {Najafabadi}}, \bibinfo {author}
  {\bibfnamefont {K.}~\bibnamefont {Watanabe}}, \bibinfo {author}
  {\bibfnamefont {T.}~\bibnamefont {Taniguchi}}, \bibinfo {author}
  {\bibfnamefont {A.}~\bibnamefont {Vishwanath}},\ and\ \bibinfo {author}
  {\bibfnamefont {P.}~\bibnamefont {Kim}},\ }\bibfield  {title} {\bibinfo
  {title} {Simularion of hubbard model physics in wse2/ws2 moire
  superlattices},\ }\href@noop {} {\bibfield  {journal} {\bibinfo  {journal}
  {Nature}\ }\textbf {\bibinfo {volume} {579}},\ \bibinfo {pages} {353}
  (\bibinfo {year} {2020})}\BibitemShut {NoStop}%
\bibitem [{\citenamefont {Park}\ \emph
  {et~al.}(2021{\natexlab{a}})\citenamefont {Park}, \citenamefont {Cao},
  \citenamefont {Watanabe}, \citenamefont {Taniguchi},\ and\ \citenamefont
  {Jarillo-Herrero}}]{ParkNat2021}%
  \BibitemOpen
  \bibfield  {author} {\bibinfo {author} {\bibfnamefont {J.~M.}\ \bibnamefont
  {Park}}, \bibinfo {author} {\bibfnamefont {Y.}~\bibnamefont {Cao}}, \bibinfo
  {author} {\bibfnamefont {K.}~\bibnamefont {Watanabe}}, \bibinfo {author}
  {\bibfnamefont {T.}~\bibnamefont {Taniguchi}},\ and\ \bibinfo {author}
  {\bibfnamefont {P.}~\bibnamefont {Jarillo-Herrero}},\ }\bibfield  {title}
  {\bibinfo {title} {Tunable strongly coupled superconductivity in magic-angle
  twisted trilayer graphene},\ }\href
  {https://doi.org/10.1038/s41586-021-03192-0} {\bibfield  {journal} {\bibinfo
  {journal} {Nature}\ }\textbf {\bibinfo {volume} {590}},\ \bibinfo {pages}
  {249} (\bibinfo {year} {2021}{\natexlab{a}})}\BibitemShut {NoStop}%
\bibitem [{\citenamefont {Hao}\ \emph {et~al.}(2021)\citenamefont {Hao},
  \citenamefont {Zimmerman}, \citenamefont {Ledwith}, \citenamefont {Khalaf},
  \citenamefont {Najafabadi}, \citenamefont {Watanabe}, \citenamefont
  {Taniguchi}, \citenamefont {Vishwanath},\ and\ \citenamefont
  {Kim}}]{HaoScience2021}%
  \BibitemOpen
  \bibfield  {author} {\bibinfo {author} {\bibfnamefont {Z.}~\bibnamefont
  {Hao}}, \bibinfo {author} {\bibfnamefont {A.}~\bibnamefont {Zimmerman}},
  \bibinfo {author} {\bibfnamefont {P.}~\bibnamefont {Ledwith}}, \bibinfo
  {author} {\bibfnamefont {E.}~\bibnamefont {Khalaf}}, \bibinfo {author}
  {\bibfnamefont {D.~H.}\ \bibnamefont {Najafabadi}}, \bibinfo {author}
  {\bibfnamefont {K.}~\bibnamefont {Watanabe}}, \bibinfo {author}
  {\bibfnamefont {T.}~\bibnamefont {Taniguchi}}, \bibinfo {author}
  {\bibfnamefont {A.}~\bibnamefont {Vishwanath}},\ and\ \bibinfo {author}
  {\bibfnamefont {P.}~\bibnamefont {Kim}},\ }\bibfield  {title} {\bibinfo
  {title} {Electric field-tunable superconductivity in alternating twist
  magic-angle trilayer graphene},\ }\href@noop {} {\bibfield  {journal}
  {\bibinfo  {journal} {Science}\ }\textbf {\bibinfo {volume} {371}},\ \bibinfo
  {pages} {1133} (\bibinfo {year} {2021})}\BibitemShut {NoStop}%
\bibitem [{\citenamefont {Zhang}\ \emph {et~al.}(2021)\citenamefont {Zhang},
  \citenamefont {Polski}, \citenamefont {Lewandowski}, \citenamefont {Thomson},
  \citenamefont {Peng}, \citenamefont {Choi}, \citenamefont {Kim},
  \citenamefont {Watanabe}, \citenamefont {Taniguchi}, \citenamefont {Alicea},
  \citenamefont {von Oppen}, \citenamefont {Refael},\ and\ \citenamefont
  {Nadj-Perge}}]{zhang2021ascendance}%
  \BibitemOpen
  \bibfield  {author} {\bibinfo {author} {\bibfnamefont {Y.}~\bibnamefont
  {Zhang}}, \bibinfo {author} {\bibfnamefont {R.}~\bibnamefont {Polski}},
  \bibinfo {author} {\bibfnamefont {C.}~\bibnamefont {Lewandowski}}, \bibinfo
  {author} {\bibfnamefont {A.}~\bibnamefont {Thomson}}, \bibinfo {author}
  {\bibfnamefont {Y.}~\bibnamefont {Peng}}, \bibinfo {author} {\bibfnamefont
  {Y.}~\bibnamefont {Choi}}, \bibinfo {author} {\bibfnamefont {H.}~\bibnamefont
  {Kim}}, \bibinfo {author} {\bibfnamefont {K.}~\bibnamefont {Watanabe}},
  \bibinfo {author} {\bibfnamefont {T.}~\bibnamefont {Taniguchi}}, \bibinfo
  {author} {\bibfnamefont {J.}~\bibnamefont {Alicea}}, \bibinfo {author}
  {\bibfnamefont {F.}~\bibnamefont {von Oppen}}, \bibinfo {author}
  {\bibfnamefont {G.}~\bibnamefont {Refael}},\ and\ \bibinfo {author}
  {\bibfnamefont {S.}~\bibnamefont {Nadj-Perge}},\ }\href@noop {} {\bibinfo
  {title} {Ascendance of superconductivity in magic-angle graphene
  multilayers}} (\bibinfo {year} {2021}),\ \Eprint
  {https://arxiv.org/abs/2112.09270} {arXiv:2112.09270 [cond-mat.supr-con]}
  \BibitemShut {NoStop}%
\bibitem [{\citenamefont {Park}\ \emph
  {et~al.}(2021{\natexlab{b}})\citenamefont {Park}, \citenamefont {Cao},
  \citenamefont {Xia}, \citenamefont {Sun}, \citenamefont {Watanabe},
  \citenamefont {Taniguchi},\ and\ \citenamefont
  {Jarillo-Herrero}}]{park2021magicangle}%
  \BibitemOpen
  \bibfield  {author} {\bibinfo {author} {\bibfnamefont {J.~M.}\ \bibnamefont
  {Park}}, \bibinfo {author} {\bibfnamefont {Y.}~\bibnamefont {Cao}}, \bibinfo
  {author} {\bibfnamefont {L.}~\bibnamefont {Xia}}, \bibinfo {author}
  {\bibfnamefont {S.}~\bibnamefont {Sun}}, \bibinfo {author} {\bibfnamefont
  {K.}~\bibnamefont {Watanabe}}, \bibinfo {author} {\bibfnamefont
  {T.}~\bibnamefont {Taniguchi}},\ and\ \bibinfo {author} {\bibfnamefont
  {P.}~\bibnamefont {Jarillo-Herrero}},\ }\href@noop {} {\bibinfo {title}
  {Magic-angle multilayer graphene: A robust family of moir\'e
  superconductors}} (\bibinfo {year} {2021}{\natexlab{b}}),\ \Eprint
  {https://arxiv.org/abs/2112.10760} {arXiv:2112.10760 [cond-mat.supr-con]}
  \BibitemShut {NoStop}%
\bibitem [{\citenamefont {Keimer}\ \emph {et~al.}(2015)\citenamefont {Keimer},
  \citenamefont {Kivelson}, \citenamefont {Norman}, \citenamefont {Uchida},\
  and\ \citenamefont {Zaanen}}]{KeimerNature2015}%
  \BibitemOpen
  \bibfield  {author} {\bibinfo {author} {\bibfnamefont {B.}~\bibnamefont
  {Keimer}}, \bibinfo {author} {\bibfnamefont {S.}~\bibnamefont {Kivelson}},
  \bibinfo {author} {\bibfnamefont {M.}~\bibnamefont {Norman}}, \bibinfo
  {author} {\bibfnamefont {S.}~\bibnamefont {Uchida}},\ and\ \bibinfo {author}
  {\bibfnamefont {J.}~\bibnamefont {Zaanen}},\ }\bibfield  {title} {\bibinfo
  {title} {From quantum matter to high-temperature superconductivity in copper
  oxides},\ }\href@noop {} {\bibfield  {journal} {\bibinfo  {journal} {Nature}\
  }\textbf {\bibinfo {volume} {518}},\ \bibinfo {pages} {179} (\bibinfo {year}
  {2015})}\BibitemShut {NoStop}%
\bibitem [{\citenamefont {Proust}\ and\ \citenamefont
  {Taillefer}(2019)}]{ProustAnnRev2019}%
  \BibitemOpen
  \bibfield  {author} {\bibinfo {author} {\bibfnamefont {C.}~\bibnamefont
  {Proust}}\ and\ \bibinfo {author} {\bibfnamefont {L.}~\bibnamefont
  {Taillefer}},\ }\bibfield  {title} {\bibinfo {title} {The remarkable
  underlying ground states of cuprate superconductors},\ }\href
  {https://doi.org/10.1146/annurev-conmatphys-031218-013210} {\bibfield
  {journal} {\bibinfo  {journal} {Annual Review of Condensed Matter Physics}\
  }\textbf {\bibinfo {volume} {10}},\ \bibinfo {pages} {409} (\bibinfo {year}
  {2019})}\BibitemShut {NoStop}%
\bibitem [{\citenamefont {Bascones}\ \emph {et~al.}(2016)\citenamefont
  {Bascones}, \citenamefont {Valenzuela},\ and\ \citenamefont
  {Calder\'on}}]{BasconesCRP2016}%
  \BibitemOpen
  \bibfield  {author} {\bibinfo {author} {\bibfnamefont {E.}~\bibnamefont
  {Bascones}}, \bibinfo {author} {\bibfnamefont {B.}~\bibnamefont
  {Valenzuela}},\ and\ \bibinfo {author} {\bibfnamefont {M.}~\bibnamefont
  {Calder\'on}},\ }\bibfield  {title} {\bibinfo {title} {Magnetic interactions
  in iron superconductors: A review},\ }\href
  {https://doi.org/https://doi.org/10.1016/j.crhy.2015.05.004} {\bibfield
  {journal} {\bibinfo  {journal} {Comptes Rendus Physique}\ }\textbf {\bibinfo
  {volume} {17}},\ \bibinfo {pages} {36 } (\bibinfo {year} {2016})}\BibitemShut
  {NoStop}%
\bibitem [{\citenamefont {Chubukov}\ and\ \citenamefont
  {Hirschfeld}(2015)}]{ChubukovPhysicsToday2015}%
  \BibitemOpen
  \bibfield  {author} {\bibinfo {author} {\bibfnamefont {A.}~\bibnamefont
  {Chubukov}}\ and\ \bibinfo {author} {\bibfnamefont {P.}~\bibnamefont
  {Hirschfeld}},\ }\bibfield  {title} {\bibinfo {title} {Iron-based
  superconductors, seven years later},\ }\href
  {https://doi.org/10.1063/PT.3.2818} {\bibfield  {journal} {\bibinfo
  {journal} {Physics Today}\ }\textbf {\bibinfo {volume} {68}},\ \bibinfo
  {pages} {46} (\bibinfo {year} {2015})}\BibitemShut {NoStop}%
\bibitem [{\citenamefont {Coleman}(2007)}]{Coleman-chapter}%
  \BibitemOpen
  \bibfield  {author} {\bibinfo {author} {\bibfnamefont {P.}~\bibnamefont
  {Coleman}},\ }\bibfield  {title} {\bibinfo {title} {Heavy fermions: electrons
  at the edge of magnetism},\ }\href@noop {} {\bibfield  {journal} {\bibinfo
  {journal} {Handbook of Magnetism and Advanced Magnetic Materials}\ }
  (\bibinfo {year} {2007})}\BibitemShut {NoStop}%
\bibitem [{\citenamefont {Si}\ and\ \citenamefont
  {Steglich}(2010)}]{SiScience2010}%
  \BibitemOpen
  \bibfield  {author} {\bibinfo {author} {\bibfnamefont {Q.}~\bibnamefont
  {Si}}\ and\ \bibinfo {author} {\bibfnamefont {F.}~\bibnamefont {Steglich}},\
  }\bibfield  {title} {\bibinfo {title} {Heavy fermions and quantum phase
  transitions},\ }\href {https://doi.org/10.1126/science.1191195} {\bibfield
  {journal} {\bibinfo  {journal} {Science}\ }\textbf {\bibinfo {volume}
  {329}},\ \bibinfo {pages} {1161} (\bibinfo {year} {2010})}\BibitemShut
  {NoStop}%
\bibitem [{\citenamefont {Guinea}\ \emph {et~al.}(2006)\citenamefont {Guinea},
  \citenamefont {Castro~Neto},\ and\ \citenamefont {Peres}}]{GuineaPRB2006}%
  \BibitemOpen
  \bibfield  {author} {\bibinfo {author} {\bibfnamefont {F.}~\bibnamefont
  {Guinea}}, \bibinfo {author} {\bibfnamefont {A.~H.}\ \bibnamefont
  {Castro~Neto}},\ and\ \bibinfo {author} {\bibfnamefont {N.~M.~R.}\
  \bibnamefont {Peres}},\ }\bibfield  {title} {\bibinfo {title} {Electronic
  states and landau levels in graphene stacks},\ }\href
  {https://doi.org/10.1103/PhysRevB.73.245426} {\bibfield  {journal} {\bibinfo
  {journal} {Phys. Rev. B}\ }\textbf {\bibinfo {volume} {73}},\ \bibinfo
  {pages} {245426} (\bibinfo {year} {2006})}\BibitemShut {NoStop}%
\bibitem [{\citenamefont {Koshino}\ and\ \citenamefont
  {McCann}(2009)}]{KoshinoPRB2009}%
  \BibitemOpen
  \bibfield  {author} {\bibinfo {author} {\bibfnamefont {M.}~\bibnamefont
  {Koshino}}\ and\ \bibinfo {author} {\bibfnamefont {E.}~\bibnamefont
  {McCann}},\ }\bibfield  {title} {\bibinfo {title} {Trigonal warping and
  berry's phase $n\ensuremath{\pi}$ in abc-stacked multilayer graphene},\
  }\href {https://doi.org/10.1103/PhysRevB.80.165409} {\bibfield  {journal}
  {\bibinfo  {journal} {Phys. Rev. B}\ }\textbf {\bibinfo {volume} {80}},\
  \bibinfo {pages} {165409} (\bibinfo {year} {2009})}\BibitemShut {NoStop}%
\bibitem [{\citenamefont {Zhang}\ \emph {et~al.}(2010)\citenamefont {Zhang},
  \citenamefont {Sahu}, \citenamefont {Min},\ and\ \citenamefont
  {MacDonald}}]{ZhangPRB2010}%
  \BibitemOpen
  \bibfield  {author} {\bibinfo {author} {\bibfnamefont {F.}~\bibnamefont
  {Zhang}}, \bibinfo {author} {\bibfnamefont {B.}~\bibnamefont {Sahu}},
  \bibinfo {author} {\bibfnamefont {H.}~\bibnamefont {Min}},\ and\ \bibinfo
  {author} {\bibfnamefont {A.~H.}\ \bibnamefont {MacDonald}},\ }\bibfield
  {title} {\bibinfo {title} {{Band structure of ABC-stacked graphene
  trilayers}},\ }\href {https://doi.org/10.1103/PhysRevB.82.035409} {\bibfield
  {journal} {\bibinfo  {journal} {Physical Review B - Condensed Matter and
  Materials Physics}\ }\textbf {\bibinfo {volume} {82}},\ \bibinfo {pages} {1}
  (\bibinfo {year} {2010})},\ \Eprint {https://arxiv.org/abs/1004.1481}
  {arXiv:1004.1481} \BibitemShut {NoStop}%
\bibitem [{\citenamefont {Zhang}\ and\ \citenamefont
  {Senthil}(2019)}]{ZhangPRB2019b}%
  \BibitemOpen
  \bibfield  {author} {\bibinfo {author} {\bibfnamefont {Y.~H.}\ \bibnamefont
  {Zhang}}\ and\ \bibinfo {author} {\bibfnamefont {T.}~\bibnamefont
  {Senthil}},\ }\bibfield  {title} {\bibinfo {title} {{Bridging Hubbard model
  physics and quantum Hall physics in trilayer graphene/h-BN moir{\'{e}}
  superlattice}},\ }\href {https://doi.org/10.1103/PhysRevB.99.205150}
  {\bibfield  {journal} {\bibinfo  {journal} {Physical Review B}\ }\textbf
  {\bibinfo {volume} {99}},\ \bibinfo {pages} {205150} (\bibinfo {year}
  {2019})},\ \Eprint {https://arxiv.org/abs/1809.05110} {arXiv:1809.05110}
  \BibitemShut {NoStop}%
\bibitem [{\citenamefont {Chittari}\ \emph {et~al.}(2019)\citenamefont
  {Chittari}, \citenamefont {Chen}, \citenamefont {Zhang}, \citenamefont
  {Wang},\ and\ \citenamefont {Jung}}]{ChittariPRL2019}%
  \BibitemOpen
  \bibfield  {author} {\bibinfo {author} {\bibfnamefont {B.~L.}\ \bibnamefont
  {Chittari}}, \bibinfo {author} {\bibfnamefont {G.}~\bibnamefont {Chen}},
  \bibinfo {author} {\bibfnamefont {Y.}~\bibnamefont {Zhang}}, \bibinfo
  {author} {\bibfnamefont {F.}~\bibnamefont {Wang}},\ and\ \bibinfo {author}
  {\bibfnamefont {J.}~\bibnamefont {Jung}},\ }\bibfield  {title} {\bibinfo
  {title} {{Gate-Tunable Topological Flat Bands in Trilayer Graphene
  Boron-Nitride Moir{\'{e}} Superlattices}},\ }\href
  {https://doi.org/10.1103/PhysRevLett.122.016401} {\bibfield  {journal}
  {\bibinfo  {journal} {Physical Review Letters}\ }\textbf {\bibinfo {volume}
  {122}},\ \bibinfo {pages} {1} (\bibinfo {year} {2019})},\ \Eprint
  {https://arxiv.org/abs/1806.00462} {arXiv:1806.00462} \BibitemShut {NoStop}%
\bibitem [{\citenamefont {Chen}\ \emph
  {et~al.}(2020{\natexlab{a}})\citenamefont {Chen}, \citenamefont {Sharpe},
  \citenamefont {Fox}, \citenamefont {Zhang}, \citenamefont {Wang},
  \citenamefont {Jiang}, \citenamefont {Lyu}, \citenamefont {Li}, \citenamefont
  {Watanabe}, \citenamefont {Taniguchi}, \citenamefont {Shi}, \citenamefont
  {Senthil}, \citenamefont {Goldhaber-Gordon}, \citenamefont {Zhang},\ and\
  \citenamefont {Wang}}]{ChenNat2020}%
  \BibitemOpen
  \bibfield  {author} {\bibinfo {author} {\bibfnamefont {G.}~\bibnamefont
  {Chen}}, \bibinfo {author} {\bibfnamefont {A.~L.}\ \bibnamefont {Sharpe}},
  \bibinfo {author} {\bibfnamefont {E.~J.}\ \bibnamefont {Fox}}, \bibinfo
  {author} {\bibfnamefont {Y.~H.}\ \bibnamefont {Zhang}}, \bibinfo {author}
  {\bibfnamefont {S.}~\bibnamefont {Wang}}, \bibinfo {author} {\bibfnamefont
  {L.}~\bibnamefont {Jiang}}, \bibinfo {author} {\bibfnamefont
  {B.}~\bibnamefont {Lyu}}, \bibinfo {author} {\bibfnamefont {H.}~\bibnamefont
  {Li}}, \bibinfo {author} {\bibfnamefont {K.}~\bibnamefont {Watanabe}},
  \bibinfo {author} {\bibfnamefont {T.}~\bibnamefont {Taniguchi}}, \bibinfo
  {author} {\bibfnamefont {Z.}~\bibnamefont {Shi}}, \bibinfo {author}
  {\bibfnamefont {T.}~\bibnamefont {Senthil}}, \bibinfo {author} {\bibfnamefont
  {D.}~\bibnamefont {Goldhaber-Gordon}}, \bibinfo {author} {\bibfnamefont
  {Y.}~\bibnamefont {Zhang}},\ and\ \bibinfo {author} {\bibfnamefont
  {F.}~\bibnamefont {Wang}},\ }\bibfield  {title} {\bibinfo {title} {{Tunable
  correlated Chern insulator and ferromagnetism in a moir{\'{e}}
  superlattice}},\ }\href {https://doi.org/10.1038/s41586-020-2049-7}
  {\bibfield  {journal} {\bibinfo  {journal} {Nature}\ }\textbf {\bibinfo
  {volume} {579}},\ \bibinfo {pages} {56} (\bibinfo {year}
  {2020}{\natexlab{a}})}\BibitemShut {NoStop}%
\bibitem [{\citenamefont {Zhou}\ \emph
  {et~al.}(2021{\natexlab{a}})\citenamefont {Zhou}, \citenamefont {Xie},
  \citenamefont {Ghazaryan}, \citenamefont {Holder}, \citenamefont {Ehrets},
  \citenamefont {Spanton}, \citenamefont {Taniguchi}, \citenamefont {Watanabe},
  \citenamefont {Berg}, \citenamefont {Serbyn},\ and\ \citenamefont
  {Young}}]{ZhouNat2021}%
  \BibitemOpen
  \bibfield  {author} {\bibinfo {author} {\bibfnamefont {H.}~\bibnamefont
  {Zhou}}, \bibinfo {author} {\bibfnamefont {T.}~\bibnamefont {Xie}}, \bibinfo
  {author} {\bibfnamefont {A.}~\bibnamefont {Ghazaryan}}, \bibinfo {author}
  {\bibfnamefont {T.}~\bibnamefont {Holder}}, \bibinfo {author} {\bibfnamefont
  {J.~R.}\ \bibnamefont {Ehrets}}, \bibinfo {author} {\bibfnamefont {E.~M.}\
  \bibnamefont {Spanton}}, \bibinfo {author} {\bibfnamefont {T.}~\bibnamefont
  {Taniguchi}}, \bibinfo {author} {\bibfnamefont {K.}~\bibnamefont {Watanabe}},
  \bibinfo {author} {\bibfnamefont {E.}~\bibnamefont {Berg}}, \bibinfo {author}
  {\bibfnamefont {M.}~\bibnamefont {Serbyn}},\ and\ \bibinfo {author}
  {\bibfnamefont {A.~F.}\ \bibnamefont {Young}},\ }\bibfield  {title} {\bibinfo
  {title} {{Half and quarter metals in rhombohedral trilayer graphene}},\
  }\href {https://doi.org/10.1038/s41586-021-03938-w} {\bibfield  {journal}
  {\bibinfo  {journal} {Nature}\ }\textbf {\bibinfo {volume} {598}},\ \bibinfo
  {pages} {429} (\bibinfo {year} {2021}{\natexlab{a}})},\ \Eprint
  {https://arxiv.org/abs/2104.00653} {arXiv:2104.00653} \BibitemShut {NoStop}%
\bibitem [{\citenamefont {Chen}\ \emph
  {et~al.}(2020{\natexlab{b}})\citenamefont {Chen}, \citenamefont {Sharpe},
  \citenamefont {Fox}, \citenamefont {Wang}, \citenamefont {Lyu}, \citenamefont
  {Jiang}, \citenamefont {Li}, \citenamefont {Watanabe}, \citenamefont
  {Taniguchi}, \citenamefont {Crommie}, \citenamefont {Kastner}, \citenamefont
  {Shi}, \citenamefont {Goldhaber-Gordon}, \citenamefont {Zhang},\ and\
  \citenamefont {Wang}}]{Chen2020}%
  \BibitemOpen
  \bibfield  {author} {\bibinfo {author} {\bibfnamefont {G.}~\bibnamefont
  {Chen}}, \bibinfo {author} {\bibfnamefont {A.~L.}\ \bibnamefont {Sharpe}},
  \bibinfo {author} {\bibfnamefont {E.~J.}\ \bibnamefont {Fox}}, \bibinfo
  {author} {\bibfnamefont {S.}~\bibnamefont {Wang}}, \bibinfo {author}
  {\bibfnamefont {B.}~\bibnamefont {Lyu}}, \bibinfo {author} {\bibfnamefont
  {L.}~\bibnamefont {Jiang}}, \bibinfo {author} {\bibfnamefont
  {H.}~\bibnamefont {Li}}, \bibinfo {author} {\bibfnamefont {K.}~\bibnamefont
  {Watanabe}}, \bibinfo {author} {\bibfnamefont {T.}~\bibnamefont {Taniguchi}},
  \bibinfo {author} {\bibfnamefont {M.~F.}\ \bibnamefont {Crommie}}, \bibinfo
  {author} {\bibfnamefont {M.~A.}\ \bibnamefont {Kastner}}, \bibinfo {author}
  {\bibfnamefont {Z.}~\bibnamefont {Shi}}, \bibinfo {author} {\bibfnamefont
  {D.}~\bibnamefont {Goldhaber-Gordon}}, \bibinfo {author} {\bibfnamefont
  {Y.}~\bibnamefont {Zhang}},\ and\ \bibinfo {author} {\bibfnamefont
  {F.}~\bibnamefont {Wang}},\ }\href {http://arxiv.org/abs/2012.10075}
  {\bibinfo {title} {{Tunable ferromagnetism at non-integer filling of a
  moir\'e superlattice}}} (\bibinfo {year} {2020}{\natexlab{b}}),\ \bibinfo
  {note} {arXiv:2012.10075}\BibitemShut {NoStop}%
\bibitem [{\citenamefont {Zhou}\ \emph
  {et~al.}(2021{\natexlab{b}})\citenamefont {Zhou}, \citenamefont {Xie},
  \citenamefont {Taniguchi}, \citenamefont {Watanabe},\ and\ \citenamefont
  {Young}}]{ZhouNat2021b}%
  \BibitemOpen
  \bibfield  {author} {\bibinfo {author} {\bibfnamefont {H.}~\bibnamefont
  {Zhou}}, \bibinfo {author} {\bibfnamefont {T.}~\bibnamefont {Xie}}, \bibinfo
  {author} {\bibfnamefont {T.}~\bibnamefont {Taniguchi}}, \bibinfo {author}
  {\bibfnamefont {K.}~\bibnamefont {Watanabe}},\ and\ \bibinfo {author}
  {\bibfnamefont {A.~F.}\ \bibnamefont {Young}},\ }\bibfield  {title} {\bibinfo
  {title} {{Superconductivity in rhombohedral trilayer graphene}},\ }\href
  {https://doi.org/10.1038/s41586-021-03926-0} {\bibfield  {journal} {\bibinfo
  {journal} {Nature}\ }\textbf {\bibinfo {volume} {598}},\ \bibinfo {pages}
  {434} (\bibinfo {year} {2021}{\natexlab{b}})},\ \Eprint
  {https://arxiv.org/abs/2106.07640} {arXiv:2106.07640} \BibitemShut {NoStop}%
\bibitem [{\citenamefont {Fazekas}()}]{Fazekas-book1998}%
  \BibitemOpen
  \bibfield  {author} {\bibinfo {author} {\bibfnamefont {P.}~\bibnamefont
  {Fazekas}},\ }\href@noop {} {\bibinfo {title} {Lecture notes on electron
  correlations and magnetism}},\ \bibinfo {note} {world Scientific Publishing
  Company}\BibitemShut {NoStop}%
\bibitem [{\citenamefont {Imada}\ \emph {et~al.}(1998)\citenamefont {Imada},
  \citenamefont {Fujimori},\ and\ \citenamefont {Tokura}}]{ImadaRMP1998}%
  \BibitemOpen
  \bibfield  {author} {\bibinfo {author} {\bibfnamefont {M.}~\bibnamefont
  {Imada}}, \bibinfo {author} {\bibfnamefont {A.}~\bibnamefont {Fujimori}},\
  and\ \bibinfo {author} {\bibfnamefont {Y.}~\bibnamefont {Tokura}},\
  }\bibfield  {title} {\bibinfo {title} {Metal-insulator transitions},\ }\href
  {https://doi.org/10.1103/RevModPhys.70.1039} {\bibfield  {journal} {\bibinfo
  {journal} {Rev. Mod. Phys.}\ }\textbf {\bibinfo {volume} {70}},\ \bibinfo
  {pages} {1039} (\bibinfo {year} {1998})}\BibitemShut {NoStop}%
\bibitem [{\citenamefont {Basov}\ \emph {et~al.}(2011)\citenamefont {Basov},
  \citenamefont {Averitt}, \citenamefont {van~der Marel}, \citenamefont
  {Dressel},\ and\ \citenamefont {Haule}}]{BasovRevModPhys2011}%
  \BibitemOpen
  \bibfield  {author} {\bibinfo {author} {\bibfnamefont {D.~N.}\ \bibnamefont
  {Basov}}, \bibinfo {author} {\bibfnamefont {R.~D.}\ \bibnamefont {Averitt}},
  \bibinfo {author} {\bibfnamefont {D.}~\bibnamefont {van~der Marel}}, \bibinfo
  {author} {\bibfnamefont {M.}~\bibnamefont {Dressel}},\ and\ \bibinfo {author}
  {\bibfnamefont {K.}~\bibnamefont {Haule}},\ }\bibfield  {title} {\bibinfo
  {title} {Electrodynamics of correlated electron materials},\ }\href
  {https://doi.org/10.1103/RevModPhys.83.471} {\bibfield  {journal} {\bibinfo
  {journal} {Rev. Mod. Phys.}\ }\textbf {\bibinfo {volume} {83}},\ \bibinfo
  {pages} {471} (\bibinfo {year} {2011})}\BibitemShut {NoStop}%
\bibitem [{\citenamefont {Calder\'on}\ and\ \citenamefont
  {Bascones}(2020)}]{CalderonPRB2020}%
  \BibitemOpen
  \bibfield  {author} {\bibinfo {author} {\bibfnamefont {M.~J.}\ \bibnamefont
  {Calder\'on}}\ and\ \bibinfo {author} {\bibfnamefont {E.}~\bibnamefont
  {Bascones}},\ }\bibfield  {title} {\bibinfo {title} {Interactions in the
  8-orbital model for twisted bilayer graphene},\ }\href
  {https://doi.org/10.1103/PhysRevB.102.155149} {\bibfield  {journal} {\bibinfo
   {journal} {Phys. Rev. B}\ }\textbf {\bibinfo {volume} {102}},\ \bibinfo
  {pages} {155149} (\bibinfo {year} {2020})}\BibitemShut {NoStop}%
\bibitem [{\citenamefont {Sch\"uler}\ \emph {et~al.}(2013)\citenamefont
  {Sch\"uler}, \citenamefont {R\"osner}, \citenamefont {Wehling}, \citenamefont
  {Lichtenstein},\ and\ \citenamefont {Katsnelson}}]{SchulerPRL2013}%
  \BibitemOpen
  \bibfield  {author} {\bibinfo {author} {\bibfnamefont {M.}~\bibnamefont
  {Sch\"uler}}, \bibinfo {author} {\bibfnamefont {M.}~\bibnamefont {R\"osner}},
  \bibinfo {author} {\bibfnamefont {T.~O.}\ \bibnamefont {Wehling}}, \bibinfo
  {author} {\bibfnamefont {A.~I.}\ \bibnamefont {Lichtenstein}},\ and\ \bibinfo
  {author} {\bibfnamefont {M.~I.}\ \bibnamefont {Katsnelson}},\ }\bibfield
  {title} {\bibinfo {title} {Optimal hubbard models for materials with nonlocal
  coulomb interactions: Graphene, silicene, and benzene},\ }\href
  {https://doi.org/10.1103/PhysRevLett.111.036601} {\bibfield  {journal}
  {\bibinfo  {journal} {Phys. Rev. Lett.}\ }\textbf {\bibinfo {volume} {111}},\
  \bibinfo {pages} {036601} (\bibinfo {year} {2013})}\BibitemShut {NoStop}%
\bibitem [{\citenamefont {Huang}\ \emph {et~al.}(2014)\citenamefont {Huang},
  \citenamefont {Ayral}, \citenamefont {Biermann},\ and\ \citenamefont
  {Werner}}]{HuangPRB2014}%
  \BibitemOpen
  \bibfield  {author} {\bibinfo {author} {\bibfnamefont {L.}~\bibnamefont
  {Huang}}, \bibinfo {author} {\bibfnamefont {T.}~\bibnamefont {Ayral}},
  \bibinfo {author} {\bibfnamefont {S.}~\bibnamefont {Biermann}},\ and\
  \bibinfo {author} {\bibfnamefont {P.}~\bibnamefont {Werner}},\ }\bibfield
  {title} {\bibinfo {title} {Extended dynamical mean-field study of the hubbard
  model with long-range interactions},\ }\href
  {https://doi.org/10.1103/PhysRevB.90.195114} {\bibfield  {journal} {\bibinfo
  {journal} {Phys. Rev. B}\ }\textbf {\bibinfo {volume} {90}},\ \bibinfo
  {pages} {195114} (\bibinfo {year} {2014})}\BibitemShut {NoStop}%
\bibitem [{\citenamefont {Georges}\ \emph {et~al.}(1996)\citenamefont
  {Georges}, \citenamefont {Kotliar}, \citenamefont {Krauth},\ and\
  \citenamefont {Rozenberg}}]{GeorgesRMP1996}%
  \BibitemOpen
  \bibfield  {author} {\bibinfo {author} {\bibfnamefont {A.}~\bibnamefont
  {Georges}}, \bibinfo {author} {\bibfnamefont {G.}~\bibnamefont {Kotliar}},
  \bibinfo {author} {\bibfnamefont {W.}~\bibnamefont {Krauth}},\ and\ \bibinfo
  {author} {\bibfnamefont {M.~J.}\ \bibnamefont {Rozenberg}},\ }\bibfield
  {title} {\bibinfo {title} {Dynamical mean-field theory of strongly correlated
  fermion systems and the limit of infinite dimensions},\ }\href
  {https://doi.org/10.1103/RevModPhys.68.13} {\bibfield  {journal} {\bibinfo
  {journal} {Rev. Mod. Phys.}\ }\textbf {\bibinfo {volume} {68}},\ \bibinfo
  {pages} {13} (\bibinfo {year} {1996})}\BibitemShut {NoStop}%
\bibitem [{\citenamefont {Kotliar}\ \emph {et~al.}(2006)\citenamefont
  {Kotliar}, \citenamefont {Savrasov}, \citenamefont {Haule}, \citenamefont
  {Oudovenko}, \citenamefont {Parcollet},\ and\ \citenamefont
  {Marianetti}}]{kotliarRMP2006}%
  \BibitemOpen
  \bibfield  {author} {\bibinfo {author} {\bibfnamefont {G.}~\bibnamefont
  {Kotliar}}, \bibinfo {author} {\bibfnamefont {S.~Y.}\ \bibnamefont
  {Savrasov}}, \bibinfo {author} {\bibfnamefont {K.}~\bibnamefont {Haule}},
  \bibinfo {author} {\bibfnamefont {V.~S.}\ \bibnamefont {Oudovenko}}, \bibinfo
  {author} {\bibfnamefont {O.}~\bibnamefont {Parcollet}},\ and\ \bibinfo
  {author} {\bibfnamefont {C.~A.}\ \bibnamefont {Marianetti}},\ }\bibfield
  {title} {\bibinfo {title} {Electronic structure calculations with dynamical
  mean-field theory},\ }\href {https://doi.org/10.1103/RevModPhys.78.865}
  {\bibfield  {journal} {\bibinfo  {journal} {Rev. Mod. Phys.}\ }\textbf
  {\bibinfo {volume} {78}},\ \bibinfo {pages} {865} (\bibinfo {year}
  {2006})}\BibitemShut {NoStop}%
\bibitem [{\citenamefont {Gull}\ \emph {et~al.}(2011)\citenamefont {Gull},
  \citenamefont {Millis}, \citenamefont {Lichtenstein}, \citenamefont
  {Rubtsov}, \citenamefont {Troyer},\ and\ \citenamefont
  {Werner}}]{GullRMP2011}%
  \BibitemOpen
  \bibfield  {author} {\bibinfo {author} {\bibfnamefont {E.}~\bibnamefont
  {Gull}}, \bibinfo {author} {\bibfnamefont {A.~J.}\ \bibnamefont {Millis}},
  \bibinfo {author} {\bibfnamefont {A.~I.}\ \bibnamefont {Lichtenstein}},
  \bibinfo {author} {\bibfnamefont {A.~N.}\ \bibnamefont {Rubtsov}}, \bibinfo
  {author} {\bibfnamefont {M.}~\bibnamefont {Troyer}},\ and\ \bibinfo {author}
  {\bibfnamefont {P.}~\bibnamefont {Werner}},\ }\bibfield  {title} {\bibinfo
  {title} {Continuous-time monte carlo methods for quantum impurity models},\
  }\href {https://doi.org/10.1103/RevModPhys.83.349} {\bibfield  {journal}
  {\bibinfo  {journal} {Rev. Mod. Phys.}\ }\textbf {\bibinfo {volume} {83}},\
  \bibinfo {pages} {349} (\bibinfo {year} {2011})}\BibitemShut {NoStop}%
\bibitem [{\citenamefont {Haule}(2007)}]{HaulePRB2007}%
  \BibitemOpen
  \bibfield  {author} {\bibinfo {author} {\bibfnamefont {K.}~\bibnamefont
  {Haule}},\ }\bibfield  {title} {\bibinfo {title} {Quantum monte carlo
  impurity solver for cluster dynamical mean-field theory and electronic
  structure calculations with adjustable cluster base},\ }\href
  {https://doi.org/10.1103/PhysRevB.75.155113} {\bibfield  {journal} {\bibinfo
  {journal} {Phys. Rev. B}\ }\textbf {\bibinfo {volume} {75}},\ \bibinfo
  {pages} {155113} (\bibinfo {year} {2007})}\BibitemShut {NoStop}%
\bibitem [{\citenamefont {Wietek}\ \emph {et~al.}(2021)\citenamefont {Wietek},
  \citenamefont {Rossi}, \citenamefont {\ifmmode~\check{S}\else
  \v{S}\fi{}imkovic}, \citenamefont {Klett}, \citenamefont {Hansmann},
  \citenamefont {Ferrero}, \citenamefont {Stoudenmire}, \citenamefont
  {Sch\"afer},\ and\ \citenamefont {Georges}}]{WietekPRX2021}%
  \BibitemOpen
  \bibfield  {author} {\bibinfo {author} {\bibfnamefont {A.}~\bibnamefont
  {Wietek}}, \bibinfo {author} {\bibfnamefont {R.}~\bibnamefont {Rossi}},
  \bibinfo {author} {\bibfnamefont {F.}~\bibnamefont {\ifmmode~\check{S}\else
  \v{S}\fi{}imkovic}}, \bibinfo {author} {\bibfnamefont {M.}~\bibnamefont
  {Klett}}, \bibinfo {author} {\bibfnamefont {P.}~\bibnamefont {Hansmann}},
  \bibinfo {author} {\bibfnamefont {M.}~\bibnamefont {Ferrero}}, \bibinfo
  {author} {\bibfnamefont {E.~M.}\ \bibnamefont {Stoudenmire}}, \bibinfo
  {author} {\bibfnamefont {T.}~\bibnamefont {Sch\"afer}},\ and\ \bibinfo
  {author} {\bibfnamefont {A.}~\bibnamefont {Georges}},\ }\bibfield  {title}
  {\bibinfo {title} {Mott insulating states with competing orders in the
  triangular lattice hubbard model},\ }\href
  {https://doi.org/10.1103/PhysRevX.11.041013} {\bibfield  {journal} {\bibinfo
  {journal} {Phys. Rev. X}\ }\textbf {\bibinfo {volume} {11}},\ \bibinfo
  {pages} {041013} (\bibinfo {year} {2021})}\BibitemShut {NoStop}%
\bibitem [{\citenamefont {Rohringer}\ \emph {et~al.}(2018)\citenamefont
  {Rohringer}, \citenamefont {Hafermann}, \citenamefont {Toschi}, \citenamefont
  {Katanin}, \citenamefont {Antipov}, \citenamefont {Katsnelson}, \citenamefont
  {Lichtenstein}, \citenamefont {Rubtsov},\ and\ \citenamefont
  {Held}}]{RohringerRMP2018}%
  \BibitemOpen
  \bibfield  {author} {\bibinfo {author} {\bibfnamefont {G.}~\bibnamefont
  {Rohringer}}, \bibinfo {author} {\bibfnamefont {H.}~\bibnamefont
  {Hafermann}}, \bibinfo {author} {\bibfnamefont {A.}~\bibnamefont {Toschi}},
  \bibinfo {author} {\bibfnamefont {A.~A.}\ \bibnamefont {Katanin}}, \bibinfo
  {author} {\bibfnamefont {A.~E.}\ \bibnamefont {Antipov}}, \bibinfo {author}
  {\bibfnamefont {M.~I.}\ \bibnamefont {Katsnelson}}, \bibinfo {author}
  {\bibfnamefont {A.~I.}\ \bibnamefont {Lichtenstein}}, \bibinfo {author}
  {\bibfnamefont {A.~N.}\ \bibnamefont {Rubtsov}},\ and\ \bibinfo {author}
  {\bibfnamefont {K.}~\bibnamefont {Held}},\ }\bibfield  {title} {\bibinfo
  {title} {Diagrammatic routes to nonlocal correlations beyond dynamical mean
  field theory},\ }\href {https://doi.org/10.1103/RevModPhys.90.025003}
  {\bibfield  {journal} {\bibinfo  {journal} {Rev. Mod. Phys.}\ }\textbf
  {\bibinfo {volume} {90}},\ \bibinfo {pages} {025003} (\bibinfo {year}
  {2018})}\BibitemShut {NoStop}%
\bibitem [{\citenamefont {Chen}\ \emph
  {et~al.}(2020{\natexlab{c}})\citenamefont {Chen}, \citenamefont {Hu},\ and\
  \citenamefont {Si}}]{Chen2020a}%
  \BibitemOpen
  \bibfield  {author} {\bibinfo {author} {\bibfnamefont {L.}~\bibnamefont
  {Chen}}, \bibinfo {author} {\bibfnamefont {H.}~\bibnamefont {Hu}},\ and\
  \bibinfo {author} {\bibfnamefont {Q.}~\bibnamefont {Si}},\ }\href
  {http://arxiv.org/abs/2007.06086} {\bibinfo {title} {{Fragile Insulator and
  Electronic Nematicity in a Graphene Moire System, arXiv:2007.06086}}}
  (\bibinfo {year} {2020}{\natexlab{c}})\BibitemShut {NoStop}%
\bibitem [{\citenamefont {Yang}\ \emph {et~al.}(2022)\citenamefont {Yang},
  \citenamefont {Chen}, \citenamefont {Han}, \citenamefont {Zhang},
  \citenamefont {Zhang}, \citenamefont {Jiang}, \citenamefont {Lyu},
  \citenamefont {Li}, \citenamefont {Watanabe}, \citenamefont {Taniguchi},
  \citenamefont {Shi}, \citenamefont {Senthil}, \citenamefont {Zhang},
  \citenamefont {Wang},\ and\ \citenamefont {Ju}}]{yang2022}%
  \BibitemOpen
  \bibfield  {author} {\bibinfo {author} {\bibfnamefont {J.}~\bibnamefont
  {Yang}}, \bibinfo {author} {\bibfnamefont {G.}~\bibnamefont {Chen}}, \bibinfo
  {author} {\bibfnamefont {T.}~\bibnamefont {Han}}, \bibinfo {author}
  {\bibfnamefont {Q.}~\bibnamefont {Zhang}}, \bibinfo {author} {\bibfnamefont
  {Y.-H.}\ \bibnamefont {Zhang}}, \bibinfo {author} {\bibfnamefont
  {L.}~\bibnamefont {Jiang}}, \bibinfo {author} {\bibfnamefont
  {B.}~\bibnamefont {Lyu}}, \bibinfo {author} {\bibfnamefont {H.}~\bibnamefont
  {Li}}, \bibinfo {author} {\bibfnamefont {K.}~\bibnamefont {Watanabe}},
  \bibinfo {author} {\bibfnamefont {T.}~\bibnamefont {Taniguchi}}, \bibinfo
  {author} {\bibfnamefont {Z.}~\bibnamefont {Shi}}, \bibinfo {author}
  {\bibfnamefont {T.}~\bibnamefont {Senthil}}, \bibinfo {author} {\bibfnamefont
  {Y.}~\bibnamefont {Zhang}}, \bibinfo {author} {\bibfnamefont
  {F.}~\bibnamefont {Wang}},\ and\ \bibinfo {author} {\bibfnamefont
  {L.}~\bibnamefont {Ju}},\ }\href@noop {} {\bibinfo {title} {Spectroscopy
  signatures of electron correlations in a trilayer graphene/hbn moir\'e
  superlattice}} (\bibinfo {year} {2022}),\ \Eprint
  {https://arxiv.org/abs/2202.12330} {arXiv:2202.12330 [cond-mat.str-el]}
  \BibitemShut {NoStop}%
\bibitem [{\citenamefont {Jarrell}\ and\ \citenamefont
  {Gubernatis}(1996)}]{JarrelPR1996}%
  \BibitemOpen
  \bibfield  {author} {\bibinfo {author} {\bibfnamefont {M.}~\bibnamefont
  {Jarrell}}\ and\ \bibinfo {author} {\bibfnamefont {J.~E.}\ \bibnamefont
  {Gubernatis}},\ }\bibfield  {title} {\bibinfo {title} {Bayesian inference and
  the analytic continuation of imaginary-time quantum monte carlo data},\
  }\href {https://doi.org/10.1016/0370-1573(95)00074-7} {\bibfield  {journal}
  {\bibinfo  {journal} {Phys. Rep.}\ }\textbf {\bibinfo {volume} {269}},\
  \bibinfo {pages} {133} (\bibinfo {year} {1996})}\BibitemShut {NoStop}%
\end{thebibliography}%
\end{document}